\documentclass[prb,amsmath,superscriptaddress,citeautoscript,twocolumn,
showpacs,floatfix]{revtex4}

\usepackage{amssymb,amsmath}
\usepackage{graphics}
\usepackage{epstopdf}
\usepackage{epsfig}
\usepackage{color}
\usepackage{amsfonts}
\usepackage{bm}
\usepackage{amscd}

\begin{document}

\title{Real-time dynamics of spin-dependent transport through a
double-quantum-dot Aharonov-Bohm interferometer with spin-orbit interaction }

\author{Matisse Wei-Yuan Tu}
\affiliation{Department of Physics, National Cheng Kung University, Tainan 70101, Taiwan}

\author{Amnon Aharony}
\email{aaharonyaa@gmail.com}
\affiliation{Physics Department, Ben Gurion University, Beer Sheva
84105, Israel} \affiliation{Raymond and Beverly Sackler School of Physics and Astronomy, Tel Aviv University,
Tel Aviv 69978, Israel}

\author{ Wei-Min Zhang}
\email{wzhang@mail.ncku.edu.tw}
\affiliation{Department of Physics, National Cheng Kung University, Tainan 70101, Taiwan}

\author{Ora Entin-Wohlman}
\affiliation{Physics Department, Ben Gurion University,  Beer Sheva
84105, Israel} \affiliation{Raymond and Beverly Sackler School of Physics and Astronomy, Tel Aviv University,
Tel Aviv 69978, Israel}

\begin{abstract}
The spin-resolved non-equilibrium real-time electron transport
through a double-quantum-dot (DQD) Aharonov-Bohm (AB) interferometer
with spin-orbit interaction (SOI) is explored.  The SOI and AB interference in the
real-time dynamics of spin transport is expressed by effective
magnetic fluxes. Analytical formulae for the time-dependent
currents, for initially unpolarized spins, are presented. In many
cases, there appear spin currents in the electrodes, for which the
spins in each electrode are polarized along characteristic
directions,
 pre-determined by the SOI
parameters and by the geometry of the system. Special choices of the
system parameters yield steady-state currents in which the spins are
fully polarized along these characteristic directions. The time
required to reach this steady state depends on the couplings of the
DQD to the leads.  The magnitudes of the currents  depend strongly
on the SOI-induced effective fluxes.  Without the magnetic flux, the
spin-polarized current cannot be sustained to the steady states, due
to the phase rigidity for this system.  For a non-degenerate DQD,
transient spin transport can be produced by the sole effects of SOI.
We also show that one can extract the spin-resolved currents from
measurements of the total charge current.
\end{abstract}



\maketitle

\section{Introduction}

Electron interference in nanoscale quantum transport systems has
long been a focus of intensive research. Of particular interest are
the Aharonov-Bohm (AB)\cite{Aharonov59485} and the Aharonov-Cahser
(AC)\cite{Aharonov84319} effects, associated with the two
fundamental degrees of freedom of an electron, namely, the charge
and the spin. By tuning externally applied fields, one is able to
modulate these interference effects and thus affect the quantum
transport properties. Interesting results of coherence modulation
have been found from the studies of stationary properties of
mesoscopic interferometer systems.
Dynamical responses of interference devices to periodically applied
driving fields have also caught attention, due to their
potential in applications. In
addition, there is a rising interest in the real time dynamics of
the charge and spin transport in such devices. This is relevant to
temporal operations of quantum devices and also to the understanding
of various physical processes. More and more attention is thus paid
to the transient evolution of coherent electron transport.
Naturally, the effects of interference on the transient dynamics of
non-equilibrium transport is an important issue. In this paper, we
study the dynamical evolution of electron transport through a
double-quantum-dot (DQD) Aharonov-Bohm interferometer with
spin-orbit interaction (SOI).

Coherence of the electron's orbital motion underlies the conductance oscillation
in the applied magnetic flux, enclosed by low-dimensional electronic systems.%
\cite{Buettiker83365,Buettiker841982,Gefen84129,Aronov87755} Studies of AB
oscillations in AB interferometers with quantum dots have been realized in experiments.\cite%
{Yacoby943149,Yacoby954047,Schuster97417,Schuster98871} Analogous to
this AB oscillation,  systems where the
SOI is present exhibit conductance
oscillations in the SOI strength, known as AC
oscillations.\cite{Mathur922964} Signatures of the AC effects have
also been
observed in experiments.\cite%
{Konig06076804,Bergsten06196803,Nagasawa12086801,Nagasawa132526}
Besides interference effects in ring-shaped structures, SOI in
nanoelectronic systems in general is known to have an important role
in spintronics.\cite{Zutic04323}  An important task in spintronics
is to generate spin-polarized currents.

An early initiative in spintronics is the proposal of the spin-field-effect
transistor by Datta and Das, that combined the SOI with ferromagnets.\cite%
{Datta90665} Optical spin injection into ferromagnets for generating
spin-polarized currents was experimentally implemented.\cite{Oiwa02137202}
Electrical spin injection from ferromagnets to semiconductors was also
realized.\cite{Jonker07542} Spin-polarized currents can also be generated
using magnetic tunnel junctions.\cite%
{LeClair02628,Santos04241203,Gajek05020406,Luders06082505} Impedance
mismatch between ferromagnets and semiconductors hinders efficient
operation of spin injection,\cite{Schmidt00R4790} whose solution
requires special techniques.\cite{Ando11655} Generating
spin-polarized currents without the use of ferromagnets, but with
tunable SOI, is an alternative option. There are two kinds of SOI in
mesoscopic electronic
structures receiving special attentions, namely, the Dresselhaus SOI\cite%
{Dresselhaus55580} and the Rashba
SOI.\cite{Rashba601224} The former is a property of  crystal
structures that lack inversion symmetry in their unit cells. The
latter, induced by the asymmetry in externally applied confinement
potential, can be controlled by tuning this external electric field.
The tunability of the Rashba SOI
strength has been demonstrated
experimentally,\cite{Konig06076804,Bergsten06196803,Nitta971335,Koga06041302}
making the utilisation of SOI for generating spin-polarized current
viable.


The simplest system that exhibits both  the AB and the AC interference
phenomena is a single loop. The loop is threaded by a magnetic flux and
an electron can flip its spins as it tunnels along the loop. By attaching
current leads to the loop, transport properties can be investigated. Spin
interference effects on the electron transport through this kind of structures
have been widely investigated. Many papers consider generating spin-polarized
currents in such systems. These cover the modulation of conductance in
one-dimensional and also two-dimensional circular rings,\cite%
{Nitta99695,Frustaglia04235310,Molnar04155335,Citro06115329} and the effects of
the coupling between the DQD and the leads on spin-dependent transport.\cite%
{Aeberhard05075328,Moldoveanu10035326} Alternative system
geometries, like polygons, have also been
studied.\cite{Bercioux05113310,Ramaglia06155328,vanVeenhuizen06235315}
Instead of using just two leads, results from attaching three leads
to the
ring, mimicking a Stern-Gerlach experiment, have also been reported.\cite%
{Foldi06155325,Chi08062106} Diamond-like loops have been found to
exhibit fully polarized spin
currents.\cite{Hatano07032107,Chen08045324,Aharony11035323} SOI in parallel DQD
with
inter-dot tunnel couplings have been considered.\cite%
{Chi07093704,Chi08343,Yin102865,Chen13035443} Though interference is
mostly effective at low temperatures, results from
a high temperature single-channel
ring are also analyzed.\cite{Shmakov13235417} Furthermore,
electron-electron
interactions have been studied in rings with SOI.\cite%
{Pletyukhov06045301,Lobos08016803} Besides focusing on the time-independent
aspect, time-periodical varying SOI has attracted attention,\cite%
{Citro06233304,Foldi09165303} as spin pumping devices.

Apart from this, much effort has been poured into the research
of time-dependent electron transport through nanojunctions.
Experimentally, time-resolved transport measurements have been implemented.%
\cite{Bylander05361,Fujisawa06759,Feve071169} Theoretically, a
multitude of approaches, focusing on many different aspects, has
been devoted to understand the
real-time electron dynamics in quantum transport.\cite%
{Cini805887,Jauho945528,Stefanucci04195318,Anders05196801,Kurth05035308,Maciejko06085324,Zheng07195127,Schmidt08235110,Muehlbacher08176403,Jin08234703,Tu08235311,Jin10083013,Segal10205323}
Real-time dynamics concerning spin-resolved currents have also been
reported. By solving the time-dependent Schr\"{o}dinger equation,
the spin-resolved time evolution of the electron wave function in a
ring with an oscillating SOI has
been analyzed.\cite{Foldi09165303} Time evolution of
the electron wave functions with
different spins has also been considered in quantum well
structures.\cite{Cruz08012004} Applying a method for
Green functions propagating in
time,\cite{Kurth05035308} transient spin-dependent currents through
a single-level dot, without a loop structure, has been
studied.\cite{Perfetto08155301}

In our previous papers,%
\cite{Tu11115318,Tu12115453} we have investigated the transient
electron dynamics in a spinless DQD AB interferometer,
based on a master equation
formalism.\cite{Tu08235311,Jin10083013} An earlier work had studied
the steady states of a similar system with spins and
SOI.\cite{Aharony11035323} Here we study the transient dynamics of
spin-dependent transport in such a system. There are clear
motivations to pursue such a study. First, spin polarization
functions are closely related to
functioning of flying spin
qubits.\cite{Aharony11035323} The dynamics of polarization processes
is thus quintessential to the processing of quantum spin information
in real time. Moreover, to coordinate the clocking of integrated
spintronic circuits, the timing of
the generation of spin-polarized currents as part of the circuit is
indispensable. Closely examining the transient spin-resolved
currents is an initiative toward these matters. Second, the ability
of the targeted system to attain full spin filtering has been proved
in the steady state.\cite{Aharony11035323} This makes it obviously
worthy to explore its spin transport dynamics. Third, the
interferometer possesses tunable coherent properties.  It is
therefore embedded with rich interference phenomena involving both
charge and spin degrees of freedom.

In this paper, we address the
following essential questions that are common to many devices
designated for the generation of spin-polarized currents, a primary
task in spintronics. These questions
are: (i) What are the factors that
determine the spin polarization directions? how do they change in
time? (ii) What are the factors that determine the magnitude of
currents of specific spins? (iii) How fast are the fully polarized
currents reached? How do we control this temporal pace? Besides all
these of operating the device, there is still an important question
in terms of basic scientific research, namely, (iv) What are
the physical mechanisms that lie
behind the answers of the above questions?

\begin{figure}[h]
\includegraphics[width=7.7cm,
height=4.0cm]{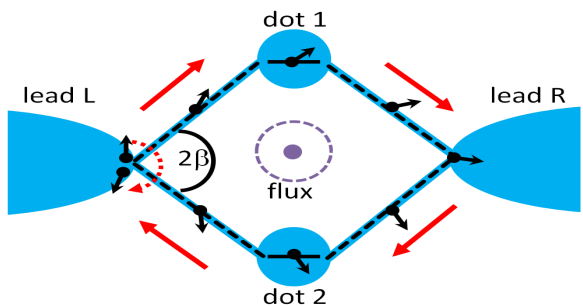} \caption{(color online) A sketch of a DQD AB
interferometer with SOI.  Each of the two dots
 contains one charge state with a spin. The dots are placed in parallel
between two leads $L$ and $R$.  A magnetic flux threads through the
loop formed by the two dots and the two leads. The electron spin
rotates as it tunnels along the loop, due to the SOI.  The angle
between the two paths is  $2\beta$.} \label{fig1}
\end{figure}

The target system in this paper is
illustrated  in Fig.~\ref{fig1}. We apply the nonequilibrium Green
function technique (NEGF) for the calculation of spin-resolved
real-time currents. We adopt the prescription for spin
transformation along electron tunneling paths given in Ref. [%
\onlinecite{Oreg922393}], and used in
[\onlinecite{Aharony11035323}]. Fully spin-polarized currents have
been obtained in the steady-state limit using the spin filter
conditions given in Ref. [\onlinecite{Aharony11035323}] (see
sections \ref{sec_realtimeDQDABSOI} and \ref{sec_SteadyState}).  The
spin-independent real-time total charge current is found to exhibit
the universal behavior pointed out in Ref. [\onlinecite{Meir89798}]
for rings with SOI (see discussions in Sec.
\ref{sec_realtimeDQDABSOI}). In addition to discussing how different
parameters of the system affect transient spin transport processes,
we also provide instructions to extract the spin-polarized currents
from the experimentally more accessible spin-independent total
charge currents.

These investigations provide brief
answers for our target system to the questions proposed above. The
polarization directions of the currents in each of the two
electrodes do not change in time and are pre-determined solely by
the SOI parameters of the system. These parameters include the
bonding geometry, adjustable in device fabrication, and the Rashba
SOI strength, controllable by the external electric field. The
applied magnetic flux, though found to be necessary in sustaining
the spin polarization of the currents in the steady-state limit,
plays no role in the determination of the polarization directions.
However, the effective fluxes, composed of the applied magnetic flux
and the SOI-induced phase, efficiently modulate the magnitudes of
the polarized currents throughout the time. The couplings between
the DQD and the electrodes then largely determine the times to reach
the final stable polarizations. These consequences can be
comprehended from the simple picture of two-path spinless
interference of the two-terminal
setup, based on the connection between the present spinful system
and its spinless counterpart.


The paper is organized as follows.
In Sec. \ref{sec_realtimeDQDABSOI},
we analytically analyze the real-time transport through a DQD AB
interferometer with SOI. In Sec.
\ref{subsec_realtimeDQDABSOI_model}, we
first introduce our model with a
description of its SOI features. In Sec.
\ref{subsec_realtimeDQDABSOI_Correspondence}, we utilize the
characteristic spinors of the SOI-induced unitary spin rotations, to
show that the Hamiltonian of the target system can be decomposed
into two commuting components, one for spin-up states and the other
for spin-down states. Based on this
decomposition, in Sec. \ref{subsec_realtimeDQDABSOI_riseSpinTrspt},
we deduce the main results about spin polarization properties
directly on the level of the Hamiltonian.  Obtaining fully
spin-polarized currents using the conditions given in Ref.
[\onlinecite{Aharony11035323}] is also shown. In Sec.
\ref{subsec_realtimeDQDABSOI_TimeDepSpinCrntGF},
the nonequilibrium formalism based
on the master equation of the density matrix of the quantum-dot
system is applied to the target system of a DQD AB interferometer
with the SOI introduced in Sec. \ref{subsec_realtimeDQDABSOI_model}.
For the purpose of tackling the dynamics purely induced by SOI, the
central area is initially prepared with no excess electrons. In this
case, the connection of the present formalism with the standard
Keldysh Green function technique is explicitly provided. To
demonstrate the functioning of
different physical factors behind
spin-polarized transport with concrete examples, we take the
commonly assumed wide-band limit for specific calculations. In Sec.
\ref{sec_SteadyState}, we first take the steady-state limit to
reassure the reproduction of fully spin-polarized currents.  We also
analyze the situation when the setup of the system deviates from
these conditions. This is followed by instructions for extracting
the spin-polarized transmission from the spin-independent total
transmission (which is much more accessible experimentally). Section
\ref{sec_DynsSpinDepTrpt} is divided into three parts.  In Sec.
\ref{subsec_DynsSpinDepTrpt_TMFullSpinPolar}, we focus on the
dynamics of getting fully spin-polarized currents.  In Sec.
\ref{subsec_DynsSpinDepTrpt_SpinCrnts}, general parameters are
explored to understand the transport of spins under the influence of
charge and spin interferences. In addition, utilizing the results
from Sec. \ref{sec_SteadyState}, we also devise similar ideas for
extracting the spin-polarized currents from the spin-independent
total charge currents in Sec.
\ref{subsec_DynsSpinDepTrpt_DeductionSpinCrnt}.  Conclusions and a
summary are given in Sec. \ref{sec_conclusion}.

\section{Real-time transport through a DQD AB interferometer with SOI}
\label{sec_realtimeDQDABSOI}


\subsection{The model}
\label{subsec_realtimeDQDABSOI_model}

The DQD interferometer which we consider here is schematically
presented in Fig.~1. It is composed of three parts, the DQD, the two
electron reservoirs on the left and on the right and the tunneling
between the DQD and the electrodes. The electron reservoirs are free
from SOI. The total Hamiltonian is then generally given by
\begin{equation}
\mathcal{H}=\mathcal{H}_{\mathrm{S}}+\mathcal{H}_{\mathrm{E}}+\mathcal{H}_{%
\mathrm{T}}\label{general_total_H}
\end{equation}

Here we focus only on SOI and ignore Zeeman splitting.  Explicitly,
the central system Hamiltonian is specified to,
\begin{equation}
\mathcal{H}_{\mathrm{S}}=\sum_{\sigma }\sum_{i=1}^{2}E_{i}a_{i\sigma
}^{\dag }a^{}_{i\sigma }.  \label{DQD_H}
\end{equation}  The DQD system is spin-degenerate with $E_{i}$'s being the on-site
energies for the single-level charge state in dot $i$. The
Hamiltonian for SOI-free\ electron reservoirs,
$\mathcal{H}_{\mathrm{E}}$, is described by
\begin{subequations}
\label{general_envH}
\begin{equation}
\mathcal{H}_{\mathrm{E}}=\sum_{\alpha }\mathcal{H}_{\alpha}
\end{equation} with
\begin{equation}
\mathcal{H}_{\alpha}=\sum_{\bm{k}\in\alpha,\sigma }\epsilon _{\alpha \bm{k}%
}c_{\alpha \bm{k}\sigma }^{\dag }c^{}_{\alpha \bm{k}\sigma}
\end{equation}
\end{subequations}where $\alpha =L,$ $R$ labels the two leads and $\bm k\sigma $ denoting the states in the leads with
orbital quantum number $\bm{k}$ and spin $\sigma $. Here $%
a_{i\sigma }^{\dagger }$ ($a_{i\sigma }^{{}}$) and $c_{\alpha
\bm{k}\sigma }^{\dag }(c^{}_{\alpha \bm{k}\sigma })$ are the
electron creation (annihilation) operators for electronic levels
$i\sigma $ and $\bm k\sigma $ in the scattering area and in the lead
$\alpha $, respectively. Due to the SOI, flipping of the spin can
occur when an electron tunnels forth and back between the DQD and
the leads.   The tunneling Hamiltonian,
\begin{equation}
\mathcal{H}_{\mathrm{T}}=\sum_{i\alpha \bm{k}}\sum_{\sigma \sigma
^{\prime
}}[V_{i\sigma ,\alpha \bm{k}\sigma ^{\prime }}a_{i\sigma }^{\dag }c^{}_{\alpha %
\bm{k}\sigma ^{\prime }}+\mathrm{h.c.}],\label{general_tulH}
\end{equation}
is then specified by the tunneling amplitudes,
\begin{equation}
V_{i\sigma ,\alpha \bm{k}\sigma ^{\prime }}=V_{i\alpha
\bm{k}}\langle \sigma |U^{i\alpha }|\sigma ^{\prime }\rangle,
\label{amplitude_decompose}
\end{equation} which contain two separate parts.
The spatial part $V_{i\alpha \bm{k}}=\bar{V}_{i\alpha
\bm{k}}e^{i\phi _{i\alpha }}$ embeds the AB phase. \ The phases are
constrained by the relation
\begin{align}
\phi =\phi _{L}-\phi _{R}, \label{AB-phase-rel}
\end{align}
with $\phi _{\alpha }=\phi _{1\alpha
}-\phi _{2\alpha }$ for $\alpha=L,R$.  Here $\phi =\Phi /\Phi _{0}$, where $%
\Phi $ is the applied magnetic flux and $\Phi _{0}$ is the flux
quantum. \ The accompanying spin rotation due to the SOI is a
unitary operation $U^{i\alpha }$, determined by the underlying
bonding geometry.

Specifically, if the system lies on the $x$-$y$ plane, then these rotations  are\cite{Aharony11035323,Oreg922393}%
\begin{subequations}
\label{Uialpha_K}
\begin{equation}
U^{i\alpha }=\exp \left( i\mathbf{K}^{i\alpha }\cdot \boldsymbol{\sigma }%
\right) \label{Uialpha_K-1}
\end{equation}%
where $\boldsymbol{\sigma }\mathbf{=}\sigma _{x}\mathbf{\hat{x}}+\sigma _{y}%
\mathbf{\hat{y}}+\sigma _{z}\mathbf{\hat{z}}$, is the vector of Pauli
matrices and%
\begin{align}
\mathbf{K}^{i\alpha }=&(\alpha
_{R}\mathbf{\hat{g}}^{i\alpha}\cdot\mathbf{\hat{y}}+\alpha
_{D}\mathbf{\hat{g}}^{i\alpha}\cdot\mathbf{\hat{x}})\mathbf{\hat{x}}\nonumber
\\&
-(\alpha _{R}\mathbf{\hat{g}}^{i\alpha}\cdot\mathbf{\hat{x}}+\alpha
_{D}\mathbf{\hat{g}}^{i\alpha}\cdot\mathbf{\hat{y}})\mathbf{\hat{y}}.
\label{K_i-alpha}
\end{align}
\end{subequations}
We denote the position of the dot $i$ by
$\mathbf{r}_{i}$ and that of the connecting site on lead $\alpha $
by $\mathbf{r}_{\alpha }$. They are separated by a distance $L$. The
unit vector pointing from dot $i$ to the connecting site on lead
$\alpha$ is then denoted by $\mathbf{\hat{g}}^{i\alpha }\mathbf{=}(\mathbf{r}%
_{\alpha }-\mathbf{r}_{i})/L$.  In Eq. (\ref{K_i-alpha}),
$\alpha _{R,D}=k_{R,D}L$ while $k_{R}$ and $k_{D}$ are the
associated coefficients for the Rashba and (linear) Dresselhaus SOI.


It is well known that an electron acquires a phase when it moves
around a loop in a region with
SOI.\cite{Aharonov84319,Oreg922393,Meir89798}\ \ In our system the
two dots and the two electrodes form a loop. \ This SOI-induced
phase is determined in the following way. \ Consider the unitary operators $U^{L}\equiv
U^{L1}U^{1R}U^{R2}U^{2L}$, $U^{1}\equiv U^{1R}U^{R2}U^{2L}U^{L1}$, $%
U^{R}\equiv U^{R2}U^{2L}U^{L1}U^{1R}$ and $U^{2}\equiv
U^{2L}U^{L1}U^{1R}U^{R2}$, where $U^{\alpha i}=\left( U^{i\alpha }\right)
^{\dag }$, which represent the rotations of the spinors related to electrons that traverse around the loop starting and ending at the
sites $L$, $1$, $R$ and $2$, respectively. \ The phase $\psi _{\text{so}}$ is
obtained by diagonalizing these spin rotations around the loop.\ \ The
results are formally given by%
\begin{equation}
U^{x}=e^{-i\psi _{\text{so}}}\left\vert \mathbf{\hat{n}}_{x}\mathbf{;+}%
\right\rangle \left\langle \mathbf{\hat{n}}_{x}\mathbf{;+}\right\vert
+e^{i\psi _{\text{so}}}\left\vert \mathbf{\hat{n}}_{x}\mathbf{;-}%
\right\rangle \left\langle \mathbf{\hat{n}}_{x}\mathbf{;-}\right\vert ,
\label{U_x_formal}
\end{equation}%
for $x=L$, $R$, $1$, and $2$. \ Here $\left\vert \mathbf{\hat{n}}_{x}\mathbf{%
;+}\right\rangle $ and $\left\vert \mathbf{\hat{n}}_{x}\mathbf{;-}%
\right\rangle $ are the spinors for spin up and spin down in the direction $%
\mathbf{\hat{n}}_{x}$ defined via $\mathbf{\hat{n}}_{x}\cdot \boldsymbol{%
\sigma }\left\vert \mathbf{\hat{n}}_{x}\mathbf{;\pm }\right\rangle =\pm
\left\vert \mathbf{\hat{n}}_{x}\mathbf{;\pm }\right\rangle $, where the $\mathbf{%
\hat{n}}_{x}$'s are certain real\  unit vectors in three dimensions.
\ The phase $\psi _{\text{so}}$ and the characteristic directions $\mathbf{%
\hat{n}}_{x}$'s are fully determined from Eq. (\ref{Uialpha_K}) and\
thus incorporate the full information about the SOI-induced spin
rotations around the loop. The authors of Ref.
[\onlinecite{Aharony11035323}] have shown that under the spin filter
conditions (see
Eq.~(\ref{cond-spinpolarcrnt}) and also discussions in Ref.
[\onlinecite{Aharony11035323}]) electrons come in\ with spinor $%
\left\vert \mathbf{\hat{n}}_{L}\mathbf{;\pm }\right\rangle $ from
the left
and will go out with spinor $\left\vert \mathbf{\hat{n}}_{R}\mathbf{;\pm }%
\right\rangle $ on the right (and vice versa).
There the explicit dependencies of
$\psi _{\text{so}}$ as well as $\mathbf{\hat{n}}_{\alpha }$ on the
bonding geometry, and on the Rashba and the Dresselhaus coefficients
have been discussed in
detail.\cite{footnote0}

\subsection{Correspondence to the spinless DQD AB interferometer}

\label{subsec_realtimeDQDABSOI_Correspondence}

\subsubsection{Decomposition into equivalent spinless systems}
\label{subsubsec_realtimeDQDABSOI_Correspondence_Decomposition}

Utilizing the eigenspinors of the rotations around the loop,
$\left\vert \mathbf{\hat{n}}_{x} ;\pm\right\rangle$, in
Eq.~(\ref{U_x_formal}), the spin rotations along the sections of the
loop, Eq.~(\ref{Uialpha_K}), become
\begin{align}
U^{i\alpha}=\sum_{\nu=\pm}e^{i\psi^{\nu}_{i\alpha}}\vert\mathbf{n}_{i};\nu\rangle\langle\mathbf{n}_{\alpha};\nu\vert.
\end{align}   Here the phases $\psi_{i\alpha }^{\pm }$'s are restrained by
\begin{equation}
\pm\psi_{\text{so}}=\psi _{L}^{\pm }-\psi _{R}^{\pm}.
\label{so_phase}
\end{equation}%
where $\psi _{\alpha }^{\pm }=\psi _{1\alpha }^{\pm }-\psi _{2\alpha
}^{\pm } $ for $\alpha =L,R$.  With the aid of the basis
transformation,
\begin{subequations}
\label{basis-trsfm}
\begin{align}
&a_{i\sigma }^{\dag }=\sum_{\nu =\pm }\left\langle \mathbf{\hat{n}}_{i}%
\mathbf{;}\nu |\sigma \right\rangle
a_{i\mathbf{\hat{n}}_{i}\mathbf{;}\nu }^{\dag },\\
&c_{\alpha %
\bm{k}\sigma }^{\dag }=\sum_{\nu =\pm }\left\langle \mathbf{\hat{n}}_{\alpha}%
\mathbf{;}\nu |\sigma \right\rangle
c_{\alpha %
\bm{k}\mathbf{\hat{n}}_{\alpha}\mathbf{;}\nu }^{\dag },
\end{align}
\end{subequations} for arbitrary spinor $\vert\sigma\rangle$, the total Hamiltonian of the system can be decomposed into two terms,
\begin{subequations}
\label{sep-H}
\begin{align}
\mathcal{H}=\mathcal{H}_{+}+\mathcal{H}_{-},
\end{align} where,
\begin{align}
\mathcal{H}_{\pm}=\mathcal{H}^{\pm}_{\mathrm{S}}+\mathcal{H}^{\pm}_{\mathrm{E}}+\mathcal{H}^{\pm}_{\mathrm{T}},
\end{align} in which
\begin{align}
\mathcal{H}^{\pm}_{\mathrm{S}}=\sum_{i=1}^{2}E_{i}a_{i
\mathbf{\hat{n}}_{i} ;\pm }^{\dag }a^{}_{i \mathbf{\hat{n}}_{i} ;\pm
},
\end{align}
\begin{align}
\mathcal{H}^{\pm}_{\mathrm{E}}=\sum_{\alpha
}\mathcal{H}^{\pm}_{\alpha},
\end{align}
\begin{align}
\mathcal{H}^{\pm}_{\alpha}=\sum_{\bm{k}\in\alpha }\epsilon _{\alpha \bm{k}%
}c_{\alpha \bm{k}\mathbf{\hat{n}}_{\alpha} ;\pm }^{\dag
}c^{}_{\alpha \bm{k}\mathbf{\hat{n}}_{\alpha} ;\pm},
\end{align}
and
\begin{align}
\mathcal{H}^{\pm}_{\mathrm{T}}=\sum_{i\alpha \bm{k}}[\bar{V}_{i\alpha \bm{k}}e^{i\varphi^{\pm}_{i\alpha}}a_{i\mathbf{\hat{n}}_{i} ;\pm }^{\dag }c^{}_{\alpha %
\bm{k}\mathbf{\hat{n}}_{\alpha} ;\pm}+\mathrm{h.c.}].
\end{align} with
\begin{align}
\varphi^{\pm}_{i\alpha}=\phi_{i\alpha}+\psi^{\pm}_{i\alpha}.
\label{effective-phases}
\end{align}
\end{subequations}  Defining similarly
$\varphi^{\pm}_{\alpha}=\varphi^{\pm}_{1\alpha}-\varphi^{\pm}_{2\alpha}$,
one directly obtains from Eqs.~(\ref{AB-phase-rel}, \ref{so_phase},
\ref{effective-phases}) that
\begin{align}
\varphi^{}_{\pm}\equiv\phi\pm\psi
_{\text{so}}=\varphi^{\pm}_{L}-\varphi^{\pm}_{R}.
\label{effective-flux}
\end{align}
The $\pm$ subscript in $\varphi^{}_{\pm}$ should not be confused with that on the operator $a^{}_{i \mathbf{\hat{n}}_{i} ;\pm
}$. The former distinguishes between the two phases in Eq. (\ref{effective-flux}), while the latter denotes the spin polarization along the dot-dependent direction $\mathbf{\hat{n}}_{i}$.

The phase relation, Eq.~(\ref{effective-flux}), in comparison to
Eq.~(\ref{AB-phase-rel}), reveals that the decomposed Hamiltonian,
$\mathcal{H}_{\pm}$,  Eq.~(\ref{sep-H}), is the
Hamiltonian for a spinless DQD AB interferometer with the flux
replaced by $\varphi^{}_{\pm}$ as the effective flux.  Furthermore, by
the orthogonality, $\langle\mathbf{\hat{n}}_{x}
;\pm\vert\mathbf{\hat{n}}_{x} ;\mp\rangle$=0, these two component
Hamiltonians commute with each other,
\begin{align}
[\mathcal{H}_{+},\mathcal{H}_{-}]=0. \label{commute-Hpm}
\end{align}
It is therefore possible to relate the spin-resolved currents for
the target system to the currents for the effective spinless setup
described by $\mathcal{H}_{+}$ and
$\mathcal{H}_{-}$ separately.

\subsubsection{Relating the spin-resolved currents to the currents for the spinless DQD AB interferometer}
\label{subsubsec_realtimeDQDABSOI_Correspondence_relationTospinless}

Consider an arbitrary spinor $\left\vert \mathbf{\hat{n};\pm }\right\rangle $%
, defined as the eigenstate of $\mathbf{\hat{n}}\cdot
\boldsymbol{\sigma }$, where $\mathbf{\hat{n}}$ is an arbitrary
three-dimensional unit vector, by
$\mathbf{\hat{n}}\cdot \boldsymbol{\sigma }\left\vert \mathbf{\hat{n};\pm }%
\right\rangle =\mathbf{\pm }\left\vert \mathbf{\hat{n};\pm
}\right\rangle $.
Taking $\vert\sigma\rangle=\left\vert \mathbf{\hat{n};\pm }\right\rangle$ in Eq.~(\ref{def-spinresolvcrnt_asdt}), the spin-resolved current on the lead $\alpha $ with the spinor $%
\left\vert \mathbf{\hat{n};\pm }\right\rangle $ is given by
\begin{equation} I_{\alpha
\mathbf{\hat{n};\pm }}\left( t\right)
=-\frac{d}{dt}\mathrm{tr}_{\mathrm{tot}}[\mathcal{N}_{\alpha,\mathbf{\hat{n};\pm
}}\rho _{\text{tot}}(t)]. \label{def_spin-dep_crnt_alpha}
\end{equation} Setting $ \mathbf{\hat{n}}=\mathbf{\hat{n}}_{\alpha}$ in
Eq.~(\ref{def_spin-dep_crnt_alpha}), with the help of the property,
Eq.~(\ref{commute-Hpm}), one is led to
\begin{subequations}
\label{spinresolv-crnt-compareto-spinless}
\begin{align}
I^{}_{\alpha \mathbf{\hat{n}}_{\alpha };\pm }\left(
t\right)=\text{tr}_{\text{tot}}\left[ \hat{I}_{\alpha
\mathbf{\hat{n}}_{\alpha }^{{}}\mathbf{;}\pm }^{{}}\left( t\right)
\rho_{\text{tot}}^{}\left( t_{0}\right) \right],
\end{align} where
\begin{align}
\hat{I}_{\alpha \mathbf{\hat{n}}_{\alpha }^{{}}\mathbf{;}\pm
}^{{}}\left( t\right)=e^{i\mathcal{H}_{\pm}(t-t_{0})}\hat{I}_{\alpha
\mathbf{\hat{n}}_{\alpha }^{{}}\mathbf{;}\pm
}^{{}}e^{-i\mathcal{H}_{\pm}(t-t_{0})},
\end{align} is the Heisenberg representation of the current
operator,
\begin{align}
\hat{I}^{}_{\alpha \mathbf{\hat{n}}_{\alpha };\pm}=-i\sum_{\bm k\in \alpha }[\bar{V}^{}_{i\alpha \bm{k}}e^{i\varphi^{\pm}_{i\alpha}}a_{i\mathbf{\hat{n}}_{i} ;\pm }^{\dag }c^{}_{\alpha %
\bm{k}\mathbf{\hat{n}}_{\alpha} ;\pm}-\mathrm{h.c.}].
\label{spinless-crnt-opt}
\end{align}
\end{subequations}

On the other hand, the current on lead $\alpha$ for the spinless
interferometer described by $\mathcal{H}_{\pm}$ with the effective
flux $\varphi^{}_{\pm}$, is defined by
\begin{align}
I^{0}_{\alpha}(\varphi^{}_{\pm},t)=-\frac{d}{dt}\mathrm{tr}_{\text{tot}}[\mathcal{N}_{\alpha,\mathbf{\hat{n}_{\alpha};\pm
}}\rho^{\pm}_{\text{tot}}(t)], \label{def-spinless-alpha-crnt-dt}
\end{align}where $\rho^{\pm}_{\text{tot}}(t)$ is the total
density matrix for the spinless system $\mathcal{H}_{\pm}$.
Similarly, Eq.~(\ref{def-spinless-alpha-crnt-dt}) can be rewritten
as
\begin{subequations}
\label{spinless-crnt-initarbi}
\begin{align}
I^{0}_{\alpha}(\varphi^{}_{\pm},t)=\text{tr}_{\text{tot}}\left[
\hat{I}^{\pm}_{\alpha }\left( t\right) \rho^{\pm}_{\text{tot}}\left(
t_{0}\right) \right], \label{def-spinless-alpha-crnt-opt}
\end{align} where
\begin{align}
\hat{I}^{\pm}_{\alpha}\left(
t\right)=e^{i\mathcal{H}_{\pm}(t-t_{0})}\hat{I}^{\pm}_{\alpha
}e^{-i\mathcal{H}_{\pm}(t-t_{0})}. \label{Heisgenberg-rep-crntopt}
\end{align}
\end{subequations}  The current operator in Eq.~(\ref{Heisgenberg-rep-crntopt}) is just
\begin{align}
\hat{I}^{\pm}_{\alpha}=\hat{I}_{\alpha \mathbf{\hat{n}}_{\alpha
}^{}\mathbf{;}\pm }^{},
\end{align} which is given by Eq.~(\ref{spinless-crnt-opt}). 

Here we want to study the spin polarization processes induced by the
intrinsic mechanisms of SOI, without the inference of the
polarization prepared in the initial states.  We hence set
$\rho_{\text{tot}}^{}\left( t_{0}\right)$  to describe an
unpolarized interferometer with the reservoirs in the thermal
equilibrium states, namely,
\begin{subequations}
\begin{align}
\rho_{\text{tot}}^{}\left(
t_{0}\right)=\rho^{}_{}(t_{0})\prod_{\alpha=L,R}\rho_{\alpha}(t_{0}),
\end{align} where
\begin{align}
\rho_{\alpha}(t_{0})=\frac{\exp\left[\left(\mathcal{H}^{}_{\alpha}-\mu_{\alpha}\mathcal{N}^{}_{\alpha}\right)/k_{B}T_{\alpha}\right]}
{\text{tr}_{\text{}}\exp\left[\left(\mathcal{H}^{}_{\alpha}-\mu_{\alpha}\mathcal{N}^{}_{\alpha}\right)/k_{B}T_{\alpha}\right]},
\end{align}
\end{subequations} and
$\mathcal{N}^{}_{\alpha}=\mathcal{N}^{}_{\alpha,\mathbf{\hat{n}_{};+
}}+\mathcal{N}^{}_{\alpha,\mathbf{\hat{n}_{};- }}$, for
an arbitrary unit vector
$\mathbf{\hat{n}}$, is the total electron number operator in lead
$\alpha$. Here $\mu_{\alpha}$ and $T_{\alpha}$ are the chemical
potential and the temperature for all spin species in lead $\alpha$.
The initial state of the DQD does not possess any polarization and
assumes the product form
$\rho(t_{0})=\rho_{\sigma}\rho_{\bar{\sigma}}$, where
$\rho_{\sigma}=\rho_{0}$ describes the state of a spinless DQD, for
all spins $\sigma$ and their opposite $\bar{\sigma}$.  Therefore one
can designate,
\begin{align}
\rho^{\pm}_{\text{tot}}(t_{0})=\rho^{}_{0}\prod_{\alpha=L,R}\frac{\exp[(\mathcal{H}^{\pm}_{\alpha}-\mu_{\alpha}\mathcal{N}^{}_{\alpha,\mathbf{\hat{n}_{\alpha};\pm
}})/k_{B}T_{\alpha}]}
{\text{tr}_{\text{}}\exp[(\mathcal{H}^{\pm}_{\alpha}-\mu_{\alpha}\mathcal{N}^{}_{\alpha,\mathbf{\hat{n}_{\alpha};\pm
}})/k_{B}T_{\alpha}]},
\end{align} to be the corresponding initial states for the effective spinless
systems, such that the following identity,
\begin{align}
I_{\alpha \mathbf{\hat{n}}_{\alpha }^{}\mathbf{;}\pm }\left(
t\right)=I^{0}_{\alpha}(\varphi^{}_{\pm},t),
\label{lead-spin-dep-crnt_form2}
\end{align} is held for all time $t$.  The identification, Eq.~(\ref{lead-spin-dep-crnt_form2}), enables us to discuss the spin-dependent transport
in the present system in terms of what has been discussed for the
spinless
 DQD AB interferometer previously.\cite{Tu12115453}  In the
steady-state limit, where the initial preparation for the part of
the DQD no longer matters, the identity
Eq.~(\ref{lead-spin-dep-crnt_form2}) with $t\rightarrow\infty$ shall
always be held.

The equality, Eq.~(\ref{lead-spin-dep-crnt_form2}), means that
$I_{\alpha }^{0}\left( \varphi^{} _{\mathbf{+}},t\right) $\ and $
I_{\alpha }^{0}\left( \varphi^{} _{\mathbf{-}},t\right) $ respectively
are the currents in lead $\alpha $ for spin-up and spin-down
electrons in the characteristic direction $\mathbf{\hat{n}}_{\alpha
}$. Using the basis transformation, Eq.~(\ref{basis-trsfm}), with
the identification, Eq.~(\ref{lead-spin-dep-crnt_form2}), the
spin-resolved current on lead $\alpha$ for an arbitrary spinor,
$\left\vert \mathbf{\hat{n};\nu }\right\rangle$, defined by
Eq.~(\ref{def_spin-dep_crnt_alpha}), can be expressed as
\begin{equation}
I_{\alpha \mathbf{\hat{n};\nu }}\left( t\right) =\sum_{\mathbf{\nu
}^{\prime
}=\pm }\left\vert \left\langle \mathbf{\hat{n}}_{\alpha }\mathbf{;\nu }%
^{\prime }|\mathbf{\hat{n};\nu }\right\rangle \right\vert
^{2}I_{\alpha }^{0}\left( \varphi^{} _{\mathbf{\nu }^{\prime
}},t\right). \label{crnt_alpha_nhat}
\end{equation}
The current formula, Eq. (\ref{crnt_alpha_nhat}), shows that $I_{\alpha \mathbf{\hat{n};\nu
}}\left( t\right)$ is a mixture of the currents $I_{\alpha
\mathbf{\hat{n}}_{\alpha };\pm }(t)$ weighted by the spinor
projections $\left\vert \left\langle \mathbf{\hat{n}}_{\alpha }\mathbf{;\pm }%
|\mathbf{\hat{n};\nu }\right\rangle \right\vert ^{2}$. \ The
arbitrary global phases embedded in $\left\vert \mathbf{\hat{n};\nu
}\right\rangle $ and $\left\vert \mathbf{\hat{n}}_{\alpha
}\mathbf{;\pm }\right\rangle $ are canceled in Eq.
(\ref{crnt_alpha_nhat}).  From either
Eq.~(\ref{lead-spin-dep-crnt_form2}) or Eq.~(\ref{crnt_alpha_nhat}),
we find that the spin-independent total current,
\begin{align}
I_{\alpha}(t)&\equiv I_{\alpha \mathbf{\hat{n};+}}\left( t\right) +I_{\alpha \mathbf{%
\hat{n};-}}\left( t\right)\notag\\
&=I_{\alpha }^{0}\left( \varphi^{} _{\mathbf{+}%
},t\right) +I_{\alpha }^{0}\left( \varphi^{}
_{\mathbf{-}},t\right),\label{totcrnt}
\end{align}is the sum
of these two currents $I_{\alpha }^{0}\left( \varphi^{}_{+},t\right) $
and $I_{\alpha }^{0}\left( \varphi^{}_{-},t\right) $.  This is
consistent with the analysis in Ref. [\onlinecite{Meir89798}].

\subsection{The rise of spin-polarized transport}
\label{subsec_realtimeDQDABSOI_riseSpinTrspt}

The main purpose of the present
paper is to explore the dynamical
 rise of the spin polarization in the currents.  This is
intimately related to the dynamics of spin flows.  The spin flow
from lead $\alpha$ is
\begin{subequations}
\label{spin-crnt-set}
\begin{equation}
\mathbf{I}_{\alpha }^{\text{S}}\left( t\right)
=-\frac{d}{dt}\text{tr}_{\text{tot}}[\mathbf{S}_{\alpha }\rho
_{\text{tot}}(t)], \label{spin-crnt-alpha}
\end{equation}%
where the total spin operator for the electrode $\alpha $ (with
$\hbar =1$) is defined by,
\begin{equation}
\mathbf{S}_{\alpha }=\sum_{\bm k\in \alpha }\sum_{\sigma\sigma^{\prime }}c_{\alpha \bm %
k\sigma}^{\dag }\left( \frac{1}{2}\boldsymbol{\sigma }\right)
_{\sigma\sigma^{\prime }}c_{\alpha \bm k\sigma^{\prime }}\text{.}
\label{spin-opt-alpha}
\end{equation}
\end{subequations}
Comparing Eq. (\ref{spin-crnt-set}) with Eq.
(\ref{def_spin-dep_crnt_alpha}), the spin flow from lead $\alpha $
is related to the spin-resolved currents there by
\begin{equation}
\mathbf{I}_{\alpha }^{\text{S}}\left( t\right) =\frac{1}{2}\sum_{i=1}^{3}%
\mathbf{\hat{x}}_{i}\left( I_{\alpha
\mathbf{\hat{x}}_{i}\mathbf{;+}}\left( t\right) -I_{\alpha
\mathbf{\hat{x}}_{i}\mathbf{;-}}\left( t\right) \right) ,
\label{alpha-spin-flow-def}
\end{equation}%
where $\left\{
\mathbf{\hat{x}}_{1},\mathbf{\hat{x}}_{2},\mathbf{\hat{x}}
_{3}\right\} =$ $\left\{
\mathbf{\hat{x}},\mathbf{\hat{y}},\mathbf{\hat{z}} \right\}$.
Using the identities $\left\vert \left\langle \mathbf{\hat{n}}%
^{\prime };\pm |\mathbf{\hat{n}};\pm \right\rangle \right\vert ^{2}=(1+%
\mathbf{\hat{n}\cdot \hat{n}}^{\prime })/2$ and $\left\vert
\left\langle \mathbf{\hat{n}}^{\prime };\pm |\mathbf{\hat{n}};\mp
\right\rangle \right\vert ^{2}=(1-\mathbf{\hat{n}\cdot
\hat{n}}^{\prime })/2$ for arbitrary directions $\mathbf{\hat{n}}$
and $\mathbf{\hat{n}}^{\prime }$ in Eq. (\ref{crnt_alpha_nhat}), we
can rewrite Eq.~(\ref{alpha-spin-flow-def}) as
\begin{equation}
\mathbf{I}_{\alpha }^{\text{S}}\left( t\right)
=\frac{\mathbf{\hat{n}}_{\alpha}}{2}\left[ I_{\alpha
\mathbf{\hat{n}}_{\alpha };+}\left( t\right) -I_{\alpha
\mathbf{\hat{n}}_{\alpha };- }\left( t\right) \right].
\label{spin-crnt-alpha-explicit}
\end{equation}  At the same time, Eq. (\ref{crnt_alpha_nhat})
becomes
\begin{equation}
I_{\alpha \mathbf{\hat{n};\pm }}\left( t\right) =\frac{I_{\alpha}(t) }{2}\pm \mathbf{I}_{\alpha }^{\text{S}}\left( t\right) \mathbf{%
\cdot \hat{n}}. \label{rewrite-spin-crnt}
\end{equation}
The factors behind
the rise of spin-polarized transports can be read from the
expression, Eq.~(\ref{rewrite-spin-crnt}).  Without the SOI, $\psi
_{\text{so}}=0$ and consequently $I_{\alpha }^{0}\left( \varphi^{}
_{\mathbf{+}},t\right) =I_{\alpha }^{0}\left(
\varphi^{} _{\mathbf{-}},t\right) =I_{\alpha }^{0}\left( \phi ,t\right) $, Eq. (%
\ref{rewrite-spin-crnt}) reduces to $I_{\alpha
\mathbf{\hat{n};+}}\left( t\right) =I_{\alpha
\mathbf{\hat{n};-}}\left( t\right) =I_{\alpha }^{0}\left( \phi
,t\right) $ for arbitrary $\mathbf{\hat{n}}$. \ Therefore without
SOI it is not possible to have spin flow in this system, as
expected. \  Only when SOI is present, the effective fluxes
$\varphi^{}_{+}$ and $\varphi^{}_{-}$ can be different.  The expression,
Eq. (\ref{rewrite-spin-crnt}), together with
Eq.~(\ref{lead-spin-dep-crnt_form2}), manifests that because spin-up
electrons and spin-down electrons experience different effective
fluxes $\varphi^{} _{+}$ and $\varphi^{} _{-}$, it is possible to have
$\mathbf{I}_{\alpha }^{\text{S}}\left( t\right) \neq 0$.  This
underlies the occurrence of a preferred spin direction in the
currents.  Note that in the steady-state limit, $t\rightarrow\infty$, the two-terminal
spinless interferometers are subjected to
phase rigidity, $I^{0}_{\alpha}(\varphi)=I^{0}_{\alpha}(-\varphi)$.
If there is no applied flux, $\phi=0$, then
$\varphi^{}_{\pm}=\pm\psi_{\text{so}}$ and therefore
$I^{0}_{\alpha}(\varphi^{}_{+})=I^{0}_{\alpha}(\varphi^{}_{-})$ and
$\mathbf{I}_{\alpha }^{\text{S}}= 0$.  This demonstrates the importance of
the combined effect of the flux and the SOI for maintaining spin
polarization in the currents to the steady-state limit.  We will
give also explicit calculations showing this result in later
sections.

A very important consequence of Eq.~(\ref{rewrite-spin-crnt}) is
that whenever there is a non-vanishing spin current
$\mathbf{I}_{\alpha }^{\text{S}}\left( t\right)\ne0$, the current on
lead $\alpha$ is always polarized in the characteristic direction
$\mathbf{\hat{n}}_{\alpha}$ for all time $t$, which is fixed by the
SOI parameters of the system.  Henceforth, to obtain a fully
spin-polarized current, one requires either $I_{\alpha
\mathbf{\hat{n}}_{\alpha };+}\left( t\right)$ or $I_{\alpha
\mathbf{\hat{n}}_{\alpha };-}\left( t\right)$ to vanish.  The
relation of Eq.~(\ref{lead-spin-dep-crnt_form2}) indicates that such
a task could be fulfilled by making one of the currents for the
effective spinless systems, $I^{0}_{\alpha}(\varphi^{}_{+},t)$
or
$I^{0}_{\alpha}(\varphi^{}_{-},t)$, diminish while the other remains
finite.  Note that since generally
$\mathbf{\hat{n}}_{L}\ne\mathbf{\hat{n}}_{R}$, the current on the
left and that on the right are polarized along different directions.

It is pointed out in Ref. [\onlinecite{Aharony11035323}] that such a
system can give rise to full spin polarization when two conditions
are fulfilled.  The first condition is that the upper arm and the
lower arm of the interferometer are symmetrically set up, namely,
\begin{subequations}
\label{cond-spinpolarcrnt}
\begin{align}
&\bar{V}_{1\alpha\bm{k}}=\bar{V}_{2\alpha\bm{k}}=\bar{V}_{\alpha\bm{k}},\label{up-dwn-equal-coup}\\
&E_{1}=E_{2}=E\label{deg}.
\end{align}  The second condition is that the applied magnetic flux and the
underlying SOI parameters should be chosen to satisfy
\begin{equation}
\cos \left( \varphi^{}_{-}\right)=-1. \label{phi_condi}
\end{equation}
\end{subequations}
These conditions were obtained from
a scattering analysis with a tight-binding modeling of the two
leads.  Indeed, applying these conditions to the total Hamiltonian
of the target system, we confirm that their validity is independent
of the energy dispersion in the leads. We also find that the rise of
the fully spin-polarized transport is equivalent to a completely
destructive interference in the corresponding spinless
interferometer, described by $\mathcal{H}_{-}$.  Such effects can be
seen by analyzing the component Hamiltonians $\mathcal{H}_{\pm}$ in
the decomposition, Eq.~(\ref{sep-H}).

To highlight the role played by the effective flux, we perform a
gauge transformation to the Hamiltonians $\mathcal{H}_{\pm}$,
yielding
\begin{subequations}
\label{gauge_trsfm_H}
\begin{align}
\mathcal{H}^{\pm}_{\mathrm{S}}=\sum_{i=1}^{2}E_{i}d_{i
\mathbf{\hat{n}}_{i} ;\pm }^{\dag }d^{}_{i \mathbf{\hat{n}}_{i} ;\pm
}, \label{gauge_trsfm_Hs}
\end{align}
\begin{align}
\mathcal{H}^{\pm}_{\mathrm{E}}=\sum_{\alpha \bm{k} }\epsilon _{\alpha \bm{k}%
}b_{\alpha \bm{k}\mathbf{\hat{n}}_{\alpha} ;\pm }^{\dag
}b^{}_{\alpha \bm{k}\mathbf{\hat{n}}_{\alpha} ;\pm},
\label{gauge_trsfm_HE}
\end{align} and
\begin{align}
&\mathcal{H}^{\pm}_{\mathrm{T}}=\notag\\
&\sum_{\bm{k}}[(\bar{\mathcal{V}}_{1L
\bm{k}}e^{i\varphi^{}_{\pm}/4}d_{1\mathbf{\hat{n}}_{1} ;\pm
}^{\dag}+\bar{\mathcal{V}}_{2L
\bm{k}}e^{-i\varphi^{}_{\pm}/4}d^{\dag}_{2\mathbf{\hat{n}}_{2};\pm
})b^{}_{L \bm{k}\mathbf{\hat{n}}_{\alpha}
;\pm}\notag\\
&+(\bar{\mathcal{V}}_{1R
\bm{k}}e^{-i\varphi^{}_{\pm}/4}d_{1\mathbf{\hat{n}}_{1} ;\pm
}^{\dag}+\bar{\mathcal{V}}_{2R
\bm{k}}e^{i\varphi^{}_{\pm}/4}d_{2\mathbf{\hat{n}}_{2} ;\pm
}^{\dag})b^{}_{R\bm{k}\mathbf{\hat{n}}_{R} ;\pm}]\notag\\
&+\mathrm{h.c.},
\label{gauge_trsfm_HT}
\end{align}
\end{subequations} where the newly defined operators and amplitudes are
\begin{subequations}
\begin{align}
&d_{1\mathbf{\hat{n}}_{1} ;\pm}^{\dag
}=e^{i\chi_{\pm}/2}a_{1\mathbf{\hat{n}}_{1} ;\pm}^{\dag }\ ,\ \ \
d_{2\mathbf{\hat{n}}_{2} ;\pm}^{\dag}=e^{-i\chi_{\pm}/2}a^{\dag }_{2\mathbf{\hat{n}}_{2};\pm}\ ,\notag\\
&b^{}_{L \bm{k}\mathbf{\hat{n}}_{L}
;\pm}=e^{i\delta\theta_{\pm}}c^{}_{L \bm{k}\mathbf{\hat{n}}_{L}
;\pm}\ ,\ \ \ b^{}_{R \bm{k}\mathbf{\hat{n}}_{R}
;\pm}=e^{-i\delta\theta_{\pm}}c^{}_{R\bm{k}\mathbf{\hat{n}}_{R}
;\pm}
\end{align} and
\begin{align}
\bar{\mathcal{V}}_{i\alpha\bm{k}}=\bar{V}_{i\alpha\bm{k}}e^{i\bar{\theta}_{\pm}},
\end{align} in which the free gauges are
\begin{align}
&\chi_{\pm}=(\varphi^{\pm}_{L}+\varphi^{\pm}_{R})/2,\notag \\
&\delta\theta_{\pm}=(\theta_{L}-\theta_{R})/4,\notag\\
&\bar{\theta}_{\pm}=(\theta_{L}+\theta_{R})/4, \label{free-gauges}
\end{align}
\end{subequations}  with
$\theta_{\alpha}=\varphi^{\pm}_{1\alpha}+\varphi^{\pm}_{2\alpha}$.

\begin{figure}[h]
\includegraphics[width=8.7cm,
height=4.0cm]{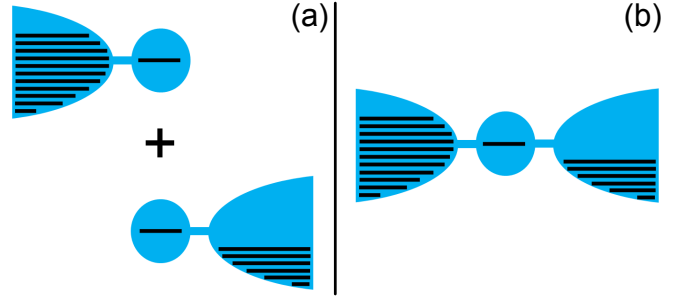} \caption{(color online) Illustration of the
effective spinless configuration mapped from the case of having
fully polarized transport current. (a) Two separate systems, each of
which is a single-level dot coupled to a reservoir.  (b) a
single-level dot coupled to two reservoirs.} \label{fig2}
\end{figure}

Applying the condition, Eq.~(\ref{up-dwn-equal-coup}), the tunneling
parts in the Hamiltonians $\mathcal{H}_{+}$ and $\mathcal{H}_{-}$,
can be written as,
\begin{subequations}
\label{tunnel-L-R-sep}
\begin{align}
\mathcal{H}^{\pm}_{\text{T}}=\mathcal{H}^{L\pm}_{\text{T}}+\mathcal{H}^{R\pm}_{\text{T}},
\end{align} where
\begin{align}
\mathcal{H}^{\alpha\pm}_{\text{T}}=
\Big[\sum_{\bm{k}\in\alpha}\bar{\mathcal{V}}_{\alpha
\bm{k}}\tilde{d}^{\dag}_{\alpha\pm}b^{}_{\alpha
\bm{k}\mathbf{\hat{n}}_{\alpha};\pm}+\text{h.c.}\Big]\sqrt{2},
\end{align} with
\begin{align}
&\tilde{d}^{\dag}_{L\pm}=\bigl(e^{i\varphi^{}_{\pm}/4}d_{1\mathbf{\hat{n}}_{1}
;\pm}^{\dag}+e^{-i\varphi^{}_{\pm}/4}d^{\dag}_{2\mathbf{\hat{n}}_{2};\pm}\bigr)/\sqrt{2},\notag\\
&\tilde{d}^{\dag}_{R\pm}=\bigl(e^{-i\varphi^{}_{\pm}/4}d_{1\mathbf{\hat{n}}_{1}
;\pm
}^{\dag}+e^{+i\varphi^{}_{\pm}/4}d^{\dag}_{2\mathbf{\hat{n}}_{2};\pm }\bigr)/\sqrt{2}\ ,
\end{align}
where the
factor $1/\sqrt{2}$ is for normalization. This shows that for $\mathcal{H}_{\pm}$ the left and the right
electrodes respectively couple to the modes $\vert
L\pm\rangle=\tilde{d}^{\dag}_{L\pm}\vert0\rangle$ and
$\vert R\pm\rangle=\tilde{d}^{\dag}_{R\pm}\vert0\rangle$,
where $\vert0\rangle$ denotes the empty state of the DQD.   The overlap between them
is
\begin{align}
\langle L\pm\vert
R\pm\rangle=\sqrt{\frac{\cos(\varphi^{}_{\pm})+1}{2}}.
\end{align}
\end{subequations}  When $\varphi^{}_{-}$ satisfies
Eq.~(\ref{phi_condi}), these two modes become orthogonal.  By
further setting the on-site energies of the DQD to be degenerate,
Eq.~(\ref{deg}), the effective spinless system described by
$\mathcal{H}_{-}$ is actually split into two separate systems, each
of which is a single-level dot coupled to a reservoir (see
Fig.~\ref{fig2}(a)).  The current on lead $\alpha$,
$I^{0}_{\alpha}(\varphi^{}_{-},t)$, in this disconnected setup will
eventually reach zero.  This picture of disconnected electrodes
underlies the completely destructive interference for the spinless
interferometer.  This effect  in turn gives
the vanishing steady-state current,
\begin{align}
I_{\alpha \mathbf{\hat{n}}_{\alpha }^{}\mathbf{;}- }\left(
t\rightarrow\infty\right)=0, \label{spin-polar-crnt-zero}
\end{align}  for the spinor
$\vert\mathbf{\hat{n}}_{\alpha }^{}\mathbf{;}-\rangle$ in lead
$\alpha$.

By the same token, the effective configuration for the connection
between the two reservoirs for $\mathcal{H}_{+}$ is controlled by
the value of $\varphi^{}_{+}$.  As long as $\varphi^{}_{+}$ does not
satisfy $\cos(\varphi^{}_{+})=-1$, the two electrodes for
$\mathcal{H}_{+}$ stay connected, supporting a non-vanishing current,
\begin{align}
I_{\alpha \mathbf{\hat{n}}_{\alpha }^{}\mathbf{;}+ }\left(
t\rightarrow\infty\right)\ne0, \label{spin-polar-crnt-nonzero}
\end{align} provided that a nonzero bias is applied.  Noticeably, when $\varphi^{}_{+}=2m\pi$ with $m$ being an arbitrary
integer, then the overlap between the two modes, $\vert L+\rangle$
and $\vert R+\rangle$ becomes unity.  This means that, at
degeneracy, the transport only goes through one mode, which is a
linear combination of the original two QD's charge states of equal
weights. This opposite limit is contrasted in Fig.~\ref{fig2}(b).

The difference between the effective configurations for
$\mathcal{H}_{-}$ and $\mathcal{H}_{+}$ has led to the different
dependencies of the dynamical evolutions of the currents
$I_{\alpha\mathbf{\hat{n}}_{\alpha};-}(t)$ and
$I_{\alpha\mathbf{\hat{n}}_{\alpha};+}(t)$ on the target system's
parameters.  The configuration of Fig.~\ref{fig2}(a) implies that
the current carrying the characteristic spinor
$\vert\mathbf{\hat{n}}_{\alpha};-\rangle$ in lead $\alpha$ is only
affected by the parameters concerning the reservoir $\alpha$ and its
coupling to the DQD, whereas the opposite reservoir $\bar{\alpha}$
exerts no influence.  On the contrary, for $\mathcal{H}_{+}$, the
connected configuration asserts that
$I_{\alpha\mathbf{\hat{n}}_{\alpha};+}(t)$ is affected by couplings
to both of the reservoirs and their respective structures,
Fig.~\ref{fig2}(b).  Explicit calculations of these spin-resolved
currents demonstrating such effects will be given in later sections.

\subsection{Real-time spin-dependent currents in terms of the Green functions}
\label{subsec_realtimeDQDABSOI_TimeDepSpinCrntGF}

In order to investigate the current polarization dynamics purely
induced by the SOI, we let the central area initially contain no
excess electrons. As we showed in
Sec. \ref{subsec_realtimeDQDABSOI_Correspondence} and Sec.
\ref{subsec_realtimeDQDABSOI_riseSpinTrspt}, the spin-dependent
currents in the target system can be discussed in terms of the
currents of the corresponding spinless systems with effective
fluxes. The transient dynamics of these spinless interferometers
have been studied before.\cite{Tu11115318,Tu12115453} To make
comparisons with our previous results easier and also to facilitate
the readers familiar with standard NEGF,\cite{Jauho945528,Haug08}
the notations of Keldysh NEGF are translated to those used in Ref.
[\onlinecite{Tu12115453}] as
\begin{subequations}
\label{gtg-gtos}
\begin{align}
\boldsymbol{\Sigma}^{r}_{\alpha}(t,\tau)&=-i{\bm g}_{\alpha}(t-\tau
),
\label{self-eng-gtos} \\
\boldsymbol{\Sigma}^{<}_{\alpha }(t,\tau)&=i\widetilde{\bm
g}_{\alpha }(t-\tau), \label{self-eng-tilde-gtos}
\end{align}
\end{subequations} for the self-energies and
\begin{subequations}
\label{uvn-gtos}
\begin{align}
\mathbf{G}^{r}(\tau,t_{0})&=-i\boldsymbol{u}(\tau ),
\label{ue-gtos}\\
\mathbf{G}^{a}(\tau,t)&=i\boldsymbol{\bar{u}}(\tau)\label{barut-gtos}, \\
\mathbf{G}^{<}(\tau,t)&=i\boldsymbol{v}(\tau ), \label{ve-gtos}
\end{align}
\end{subequations} for the Green functions.\cite{Jin10083013} The
spin-resolved real-time current in terms of Keldysh NEGF is
summarized in Appendix \ref{sec_formalism}.


The current for the effective spinless interferometer
reads\cite{Tu12115453}
\begin{align}
&I_{\alpha }^{0}\left( \varphi^{} _{\mathbf{\pm }},t\right)  \notag \\
&=2\text{ReTr}\int_{0}^{t}d\tau \left( \mathbf{\tilde{g}}_{\alpha
}^{\pm
}\left( t-\tau \right) \mathbf{\bar{u}}^{\pm }\left( \tau \right) -\mathbf{g}%
_{\alpha }^{\pm }\left( t-\tau \right) \mathbf{v}^{\pm }\left( \tau \right)
\right).  \notag \\
&  \label{crnt_alpha_plusminus}
\end{align} The Green functions for the effective spinless system satisfy the
equations,
\begin{subequations}
\label{GF-u-v-set}
\begin{equation}
\partial _{\tau }\mathbf{u}^{\pm }\left( \tau \right) +i\mathbf{Eu}^{\pm
}\left( \tau \right) +\int_{t_{0}}^{\tau }d\tau ^{\prime
}\mathbf{g}^{\pm }\left( \tau -\tau ^{\prime }\right)
\mathbf{u}^{\pm }\left( \tau ^{\prime }\right) =0,\label{u_pol_comp}
\end{equation}%
and%
\begin{equation}
\mathbf{v}^{\pm }\left( \tau \right) =\int_{t_{0}}^{\tau }d\tau
_{1}\int_{t_{0}}^{t}d\tau _{2}\mathbf{u}^{\pm }\left( \tau -\tau
_{1}\right)
\mathbf{\tilde{g}}^{\pm }\left( \tau _{1}-\tau _{2}\right) \mathbf{\bar{u}}%
^{\pm }\left( \tau _{2}\right) ,\label{v_pol_comp}
\end{equation}
\end{subequations} with $\mathbf{\bar{u}}^{\pm }\left( \tau \right) =\mathbf{u}^{\pm }\left(
t-\tau +t_{0}\right) ^{\dag }$ and the boundary condition
$\mathbf{u}^{\pm }\left( 0\right) =\mathbf{1}$ is imposed. The
self-energies for the effective spinless system are $\mathbf{g}^{\pm
}\left( \tau \right) =\sum_{\alpha }\mathbf{g}_{\alpha }^{\pm
}\left( \tau \right) $ and $\mathbf{\tilde{g}}^{\pm }\left( \tau
\right) =\sum_{\alpha }\mathbf{\tilde{g}}_{\alpha }^{\pm }\left(
\tau \right) $ with
\begin{subequations}
\begin{align}
&\left[ \mathbf{g}_{\alpha }^{\pm }\left( \tau \right) \right]
_{ij}= \left[ \mathbf{\bar{g}}_{\alpha }\left( \tau \right) \right]
_{ij}e^{i\left( \varphi _{i\alpha }^{\pm }-\varphi_{j\alpha }^{\pm
}\right) },\\& \left[ \mathbf{\tilde{g}}_{\alpha }^{\pm }\left( \tau
\right) \right] _{ij}=\left[ \mathbf{\tilde{\bar{g}}}_{\alpha
}\left( \tau \right) \right] _{ij}e^{i\left( \varphi_{i\alpha }^{\pm
}-\varphi_{j\alpha }^{\pm }\right) },
\end{align} and
\begin{align}
&\bar{\mathbf{g}}_{\alpha }\left( \tau \right) =\int \frac{d\omega
}{2\pi }\mathbf{\bar{\Gamma}}^{\alpha }\left( \omega \right)
e^{-i\omega \tau },\\ &
\mathbf{\tilde{%
\bar{g}}}_{\alpha }\left( \tau \right) =\int_{-\infty }^{\infty }\frac{%
d\omega }{2\pi }f_{\alpha }\left( \omega \right) \mathbf{\bar{\Gamma}}%
^{\alpha }\left( \omega \right) e^{-i\omega \tau },
\end{align} in which
\begin{equation}
\left[ \mathbf{\bar{\Gamma}}^{\alpha }\left( \omega \right) \right]
_{ij}=2\pi\sum_{k\in \alpha }\bar{V}_{i\alpha k}\bar{V}_{j\alpha
k}^{\ast }\delta \left( \omega -\varepsilon _{\alpha k}\right)
\text{.} \label{lv-brd_space_noAB}
\end{equation}
\end{subequations}
Here $\mathbf{E}=\left(
\begin{array}{cc}
E_{1} & 0 \\
0 & E_{2}%
\end{array}%
\right) $ is the on-site energy matrix for the DQD.  Explicitly,
the matrices of the effective self-energies are,%
\begin{widetext}
\begin{equation}
\mathbf{g}^{\pm }\left( \tau \right) =\left(
\begin{array}{cc}
\left[ \mathbf{\bar{g}}_{L}\left( \tau \right) \right] _{11}+\left[ \mathbf{%
\bar{g}}_{R}\left( \tau \right) \right] _{11} & e^{i\chi _{\pm
}}\left( \left[ \mathbf{\bar{g}}_{L}\left( \tau \right) \right]
_{12}e^{i\varphi^{} _{\pm }/2}+\left[ \mathbf{\bar{g}}_{R}\left(
\tau \right) \right]
_{12}e^{-i\varphi^{} _{\pm }/2}\right)  \\
e^{-i\chi _{\pm }}\left( \left[ \mathbf{\bar{g}}_{L}\left( \tau \right) %
\right] _{21}e^{-i\varphi^{} _{\pm }/2}+\left[
\mathbf{\bar{g}}_{R}\left( \tau
\right) \right] _{21}e^{i\varphi^{} _{\pm }/2}\right)  & \left[ \mathbf{\bar{g}}%
_{L}\left( \tau \right) \right] _{22}+\left[
\mathbf{\bar{g}}_{R}\left( \tau
\right) \right] _{22}%
\end{array}%
\right) ,\label{effective_self-eng}
\end{equation}
\end{widetext}where
$\chi _{\pm }$ is an arbitrary gauge phase given in
Eq.~(\ref{free-gauges}). Straightforwardly, solving
Eq.~(\ref{u_pol_comp}) by Laplace transformation and substituting
the solution into Eq.~(\ref{crnt_alpha_plusminus}), we found the
current, as a physical observable, is independent of the arbitrary
gauge phase $\chi _{\pm }$ that appears in Eq.
(\ref{effective_self-eng}). By taking
$\mathbf{\hat{n}=\hat{n}}_{\alpha }$ in Eq.~(\ref{crnt_alpha_nhat}),
together with Eq.~(\ref{crnt_alpha_plusminus}), one immediately
verifies Eq.~(\ref{lead-spin-dep-crnt_form2}).

The Green functions, Eq.~(\ref{GF-u-v-set}), together with the
current expression, Eq.~(\ref{crnt_alpha_plusminus}), and the
identity, Eq.~(\ref{lead-spin-dep-crnt_form2}), form the basis of
exploring the spin-dependent electron transport described by Eq.
(\ref{rewrite-spin-crnt}). The self-energies
$\mathbf{g}^{\pm}(\tau)$, $\mathbf{\tilde{g}}^{\pm}(\tau)$ and the
Green functions $\mathbf{u}^{\pm }(\tau)$, $\mathbf{v}^{\pm}(\tau)$
are respectively $\boldsymbol{g}(\tau)$,
$\boldsymbol{\tilde{g}}(\tau)$ and $\boldsymbol{u}(\tau)$,
$\boldsymbol{v}(\tau)$ found through the replacement of $\phi$ in
Ref. [\onlinecite{Tu12115453}] by $\varphi^{}_{\pm}$.

For explicit calculations, we take the commonly assumed wide-band
limit.  The effective self-energy functions,
Eq.(\ref{effective_self-eng}), become
\begin{align}
\mathbf{g}^{\pm }\left( \tau \right)=\delta \left( \tau
\right)\left(
\begin{array}{cc}
\Gamma & e^{i\chi _{\pm }}m_{\pm } \\
e^{-i\chi _{\pm }}m^{*}_{\pm } & \Gamma%
\end{array}%
\right),  \label{g_plus_minus}
\end{align} where $m_{\pm }=\left( \Gamma _{L}e^{i\varphi^{}
_{\pm }/2}+\Gamma _{R}e^{-i\varphi^{} _{\pm }/2}\right)$. The
broadening due to the coupling to electrode $\alpha $ is $\Gamma
_{\alpha } $ and $\Gamma =\Gamma _{L}+\Gamma _{R}$. The explicit
expressions of the Green functions as well as the spinless currents
under the wide-band assumption can be found in Ref.
[\onlinecite{Tu12115453}].

\section{ Spin-dependent current in the steady state}
\label{sec_SteadyState}

Before we proceed to investigate the dynamical processes, we first
take the steady-state limit. \ We examine the conditions for
generating spin-polarized current. \ We  reproduce
known results about spin polarization. \ We also discuss other
possibilities of spin dependence of the steady-state currents.
In the end of this section, we
investigate how to extract the spin-polarized transmission from the
spin-independent total transmission at various electric and magnetic
fields.

\subsection{Spin-polarized currents
in the steady state}

\label{subsec_SteadyState_SpinPolar}

The spin-resolved current, $I_{\alpha\mathbf{\hat{n}};\pm }(t)$,
along an arbitrary direction, $\mathbf{n}$, involves both the
currents $I^{0}_{\alpha}(\varphi^{}_{+},t)$ and
$I^{0}_{\alpha}(\varphi^{}_{-},t)$. As seen from Ref.
[\onlinecite{Tu12115453}], each of these currents may require a
different time to approach its steady state, determined from the
rates,
\begin{subequations}
\begin{align} \gamma _{\varphi^{}
_{\pm }}^{\pm }=\frac{1}{2}\left( \Gamma \pm \Gamma _{\varphi^{}
_{\pm }}\right),
\end{align} depending on the effective flux, $\varphi^{}_{\pm}$, through
\begin{align}
\Gamma _{\varphi^{} _{\pm }}=\sqrt{\Gamma _{L}^{2}+\Gamma
_{R}^{2}+2\Gamma _{L}\Gamma _{R}\cos \left( \varphi^{} _{\pm
}\right) -\delta E^{2}}, \label{Gamma_phi}
\end{align}
\end{subequations} where $\delta{E}=E_{1}-E_{2}$.
When $\gamma _{\varphi^{} _{\pm }}^{-}$ is nonzero, the time for
$I^{0}_{\alpha}(\varphi^{}_{\pm},t)$ to reach its steady state must
be larger than $ 1/\text{Re}[\gamma _{\varphi^{} _{\pm }}^{-}]$.
When $\delta{E}=0$ and the effective flux $\varphi^{}_{\pm}$ is an
even multiple of $\pi$,  then $\gamma^{-}_{\varphi^{}_{\pm}}=0$ and
$\gamma^{+}_{\varphi^{}_{\pm}}=\Gamma$.  In this case, as discussed
previously,\cite{Tu12115453} only a single decay channel is present
and the corresponding decay rate is just $\Gamma$.  The time to
reach the steady state in this situation must be lager than
$1/\Gamma$.

Taking the steady-state limit of Eq. (\ref{rewrite-spin-crnt}),
$I_{\alpha\mathbf{\hat{n};\pm
}}(t\rightarrow\infty)=I^\infty_{\alpha\mathbf{\hat{n};\pm
}}$, the current from the lead $\alpha$ carrying spinor
$|\mathbf{\hat{n}};\pm\rangle$ becomes
\begin{subequations}
\label{steady-state-current}
\begin{equation}
I^\infty_{\alpha \mathbf{\hat{n}\pm }}=\frac{1}{2}\left[ \left(
1\pm\mathbf{\hat{n}}%
_{\alpha }\mathbf{\cdot \hat{n}}\right) I_{\alpha }^{0}\left(
\varphi^{}
_{\mathbf{+}}\right) +\left( 1\mp\mathbf{\hat{n}}%
_{\alpha }\mathbf{\cdot \hat{n}}%
\right) I_{\alpha }^{0}\left( \varphi^{} _{\mathbf{-}}\right) \right]
\label{steady-crnt}
\end{equation}%
where%
\begin{equation}
I_{\alpha }^{0}\left( \varphi^{} _{\pm}\right) =\int \frac{d\omega
}{2\pi }\left( f_{\alpha }\left( \omega \right)
-f_{\bar{\alpha}}\left( \omega \right) \right) T^{0}\left(\varphi^{}
_{\pm },\omega \right) , \label{effective-steady-crnt}
\end{equation}in which $\bar{\alpha}=R$
if $\alpha =L$ and vice versa.  Here
the linear response of the effective spinless\ system with
effective flux $\varphi^{} _{\pm }$ is given by the transmission,
\begin{equation}
T^{0}\left(\varphi^{} _{\pm }, \omega \right) =4\Gamma _{L}\Gamma
_{R}\frac{\omega
^{2}\cos ^{2}\left( \varphi^{} _{\mathbf{\pm }}/2\right) +\left( \frac{\delta E%
}{2}\sin \left( \varphi^{} _{\mathbf{\pm }}/2\right) \right) ^{2}}{\left(
\omega ^{2}+\left( \gamma _{\varphi^{} _{\pm }}^{+}\right) ^{2}\right) \left(
\omega ^{2}+\left( \gamma _{\varphi^{} _{\pm }}^{-}\right) ^{2}\right) }\text{.}
\label{transmission_plus_minus}
\end{equation}
\end{subequations}
Due to charge conservation, the steady-state currents for the
spinless DQD AB interferometer are subjected to
\begin{align}
I^{0}_{L}(\varphi^{}_{\pm})=-I^{0}_{R}(\varphi^{}_{\pm}).
\end{align}
 Using Eq. (\ref{lead-spin-dep-crnt_form2}), we
immediately find that
\begin{align}
I^\infty_{L \mathbf{\hat{n}}_{L}\pm }=-I^\infty_{R \mathbf{\hat{n}}_{R}\pm }.
\label{spin-rot-crnt-conserv}
\end{align}  This indicates that the current for spinor $\left\vert
\mathbf{\hat{n}}_{L};\pm \right\rangle $ leaving the left side is
converted to the current for $\left\vert \mathbf{\hat{n}}_{R};\pm
\right\rangle $ in the right side.  It also reveals the effects of SOI
when electrons are transferred across the DQD from one lead to the
other. Directly from Eq. (\ref{spin-rot-crnt-conserv}) or more
generally from Eq. (\ref{steady-state-current}), we have the total
current conversation,
\begin{align}
I_{L}=-I_{R}. \label{crnt-conserv}
\end{align}

Applying the conditions for realizing the full spin-polarization in
Eq.~(\ref{cond-spinpolarcrnt}) to Eq. (\ref{steady-state-current})
results in
\begin{subequations}
\label{spin-polar-crnt-wb}
\begin{align}
I^\infty_{\alpha \mathbf{\hat{n}_{\alpha};- }}=0,\\I^\infty_{\alpha
\mathbf{\hat{n}_{\alpha};+ }}\ne0.
\end{align}
\end{subequations} Therefore the current on lead $\alpha$ is
polarized to carry only the spinor $\left\vert
\mathbf{\hat{n}}_{\alpha};+ \right\rangle $ while that for the
opposite spinor $\left\vert \mathbf{\hat{n}}_{\alpha};-
\right\rangle $ vanishes.   Equations~(\ref{spin-polar-crnt-wb}),
obtained in the wide-band limit, are the same as
Eqs.~(\ref{spin-polar-crnt-zero},\ref{spin-polar-crnt-nonzero}),
deduced independently of the form of the level-broadening function
given in Sec. \ref{subsec_realtimeDQDABSOI_riseSpinTrspt}.  Our
results here thus reproduce the findings in Ref.
[\onlinecite{Aharony11035323}].

Given that the two conditions for full polarization are satisfied,
the polarized current on lead $\alpha$ is described by
\begin{align}
I^\infty_{\alpha \mathbf{\hat{n}_{\alpha};+ }}=4\Gamma _{L}\Gamma _{R}\int
\frac{d\omega }{2\pi }\left( f_{\alpha }\left( \omega \right)
-f_{\bar{\alpha}}\left( \omega \right)
\right)\notag\\\times\frac{\omega ^{2}\cos ^{2}\left( \varphi^{} _{+
}/2\right) }{\left( \omega ^{2}+\left( \gamma _{\varphi^{} _{+
}}^{+}\right) ^{2}\right) \left( \omega ^{2}+\left( \gamma _{\varphi^{}
_{+ }}^{-}\right) ^{2}\right) }.\label{steady-alpha-plus-crnt}
\end{align}
Setting
$\mu_{L}=-\mu_R=eV/2$ at zero temperature, the integral in the above
equation can be done explicitly, yielding
\begin{align}
&I^\infty_{\alpha \mathbf{\hat{n}_{\alpha};+ }}= \pm\frac{4\Gamma_L\Gamma_R}{\pi\Gamma\Gamma_\varphi}\cos ^{2}
\left( \varphi^{} _{\mathbf{+}}/2\right) \notag \\
&\times\Bigl[ \gamma _{\varphi^{} _{+}}^{+}\tan ^{-1}\left( \frac{eV}{%
2\gamma _{\varphi^{} _{+}}^{+}}\right)- \gamma _{\varphi^{} _{+}}^{-}\tan ^{-1}\left( \frac{eV}{%
2\gamma _{\varphi^{} _{+}}^{-}}\right)
 \Bigr]\ ,
\label{polarized_trps_steady}
\end{align}
where the overall sign $\pm$ on the right-hand side
 takes $+$ for $\alpha=L$ and
$-$ for $\alpha=R$. This shows that
the magnitude of the fully polarized currents sensitively
depends on the effective flux
$\varphi^{} _{\mathbf{+}}$ mainly through the term $\cos ^{2}\left(
\varphi^{} _{\mathbf{+}}/2\right) $.


Opposite to the fully spin-polarized current is the randomly
polarized current, namely,
$I_{\alpha\mathbf{\hat{n};+}}=I_{\alpha\mathbf{\hat{n};-}}$, for any
direction $\mathbf{\hat{n}}$.  There are two possibilities for such
unpolarized transport to occur.  The first is the trivial situation
where the SOI is switched off.  The currents remain unpolarized not
only in the steady-state limit but also throughout the time, as
shown in the previous discussion below Eq.
(\ref{rewrite-spin-crnt}).  The second circumstance is that there is
no applied magnetic flux, $\phi =0$.  In this case, we have $\varphi^{}
_{\pm }=\pm \psi _{\text{so}}$ and $I_{\alpha
}^{0}\left( \psi _{\text{so}}\right) =I_{\alpha}^{0}\left( -\psi _{\text{%
so}}\right) $ due to phase rigidity because the effective spinless
interferometer is a two-terminal system [see also Eqs.
(\ref{effective-steady-crnt},\ref{transmission_plus_minus})].
Putting this into Eq.~(\ref{steady-crnt}), one immediately obtains
$I_{\alpha\mathbf{\hat{n};+}}=I_{\alpha\mathbf{\hat{n};-}}$, for all
directions $\mathbf{\hat{n}}$.  This result  exemplifies the
discussion about the steady-state limit below
Eq.~(\ref{rewrite-spin-crnt}) in Sec.
\ref{subsec_realtimeDQDABSOI_riseSpinTrspt}.  The currents in the
steady-state limit are unpolarized but they may be transiently
polarized (see the later discussion in Sec.
\ref{subsec_DynsSpinDepTrpt_SpinCrnts}). When one is away from
either of the two conditions for full spin polarization, one cannot
reach $T^{0}\left(\varphi^{}_{-}, \omega \right) =0$\ for all
$\omega $. Therefore in general both $I_{\alpha
\mathbf{\hat{n}_{\alpha}- }}$ and $I_{\alpha
\mathbf{\hat{n}_{\alpha}+ }}$ are nonzero and the currents are only
partially polarized.


\subsection{Extracting spin-resolved
transmission from measurement of total transmission}

\label{subsec_SteadyState_ExtractSpTrsm}

In the absence of the magnetic flux, the line shape of the
spin-independent total transmission depends solely on the
SOI-induced phase.  Therefore one can extract $\psi_{\text{so}}$
from the total transmission.  This in turn can be used to map out
the relations between the values of the transmission for the
underlying spinless system and the effective fluxes,
$T^{0}(\varphi,\omega)$.

The spin-independent total charge
transport current is defined by
\begin{align}
I=\frac{1}{2}(I_{L}-I_{R}).\label{steady-totcrnt}
\end{align}  Using Eq.~(\ref{steady-state-current}) with Eq.~(\ref{totcrnt}), the charge transport current, Eq.~(\ref{steady-totcrnt}), in the steady-state
limit becomes
\begin{subequations}
\label{tot_chgcrnt-stdy}
\begin{align}
I=\int \frac{d\omega }{2\pi }\left( f_{L }\left( \omega \right)
-f_{R}\left( \omega \right) \right)
T\left(\phi,\psi_{\text{so}},\omega \right),
\end{align} where the total charge transmission, $T\left(\phi,\psi_{\text{so}},\omega
\right)$, is
\begin{align}
T\left(\phi,\psi_{\text{so}},\omega
\right)=T^{0}\left(\varphi^{}_{+},\omega
\right)+T^{0}\left(\varphi^{}_{-},\omega \right).
\end{align}
\end{subequations}
The zero bias charge conductance at low temperature, as the linear response,  is simply given by $T\left(\phi,\psi_{\text{so}},\omega
\right)$ in Eq.~(\ref{tot_chgcrnt-stdy}).  By turning off the
magnetic flux, the total charge transmission becomes
\begin{align}
T\left(\phi=0,\psi_{\text{so}},\omega
\right)=2T^{0}\left(\psi_{\text{so}},\omega \right),
\end{align} where we have utilized $T^{0}(\psi_{\text{so}},\omega
)=T^{0}(-\psi_{\text{so}},\omega )$.

\begin{figure}[h]
\includegraphics[width=9.0cm,
height=5.5cm]{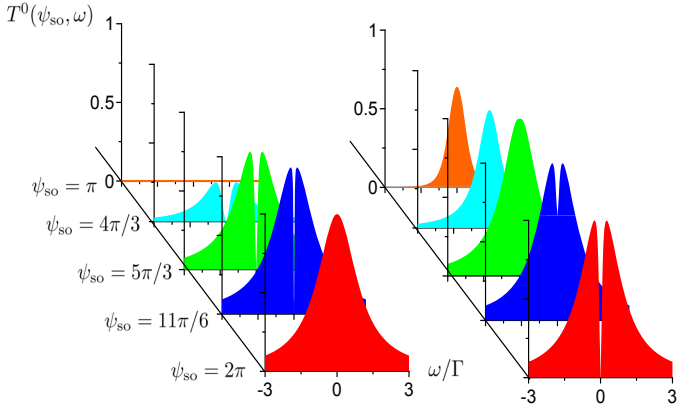} \caption{(color online) The transmission for
the effective spinless DQD AB interferometer with the flux set at
$\varphi=\psi_{\text{so}}$.  The left panel is for $\delta{E}=0$ and
the right panel is for $\delta{E}=0.5\Gamma$.  The couplings are
$\Gamma_{L}=\Gamma_{R}=\Gamma/2$ for both panels.  These series of
stacked plots show $T^{0}(\psi_{\text{so}},\omega)$ versus $\omega$
at some specific values of $\psi_{\text{so}}$, as labeled in the
figure.} \label{fig2-1}
\end{figure}

We discuss separately the cases with
$\delta{E}=0$ and with $\delta{E}\ne0$.
By tuning the on-site energies such
that $\delta{E}=0$, with $\psi_{\text{so}}=2n\pi$, where $n$ is an
arbitrary integer, Eq.~(\ref{transmission_plus_minus}) becomes a
Lorentzian line shape,
\begin{align}
T^{0}\left(2n\pi,\omega \right) =\frac{4\Gamma _{L}\Gamma
_{R}}{\left( \omega ^{2}+\Gamma ^{2}\right)
},\label{lorentz-lineshape}
\end{align} as shown by the most front plot in the left panel
of Fig.~\ref{fig2-1}.  For $\psi_{\text{so}}\ne2n\pi$, the
single-peak profile  splits into two peaks, as shown by the
other plots in the left panel of Fig.~ \ref{fig2-1}.  The separation
of the two peaks is given by
\begin{align}
\Delta{\omega}=\sqrt{2\Gamma_{L}\Gamma_{R}(1-\cos(\psi_{\text{so}}))}.\label{peak-distance}
\end{align}  For $\psi_{\text{so}}=(2n-1)\pi$,  the
transmissions vanish, as shown by the most rear plot on the left
panel of Fig.~\ref{fig2-1}.

With $\delta{E}\ne0$, $T^{0}\left(\psi_{\text{so}},\omega \right)$
for $\psi_{\text{so}}=2n\pi$ shows split peaks and its value at
$\omega=0$ equals to zero.  This is the most front plot on the right
panel of Fig.~\ref{fig2-1}.  The separation between the two peaks in
this case becomes
\begin{align}
\Delta{\omega}=\sqrt{2}|\delta{E}|.\label{peak-distance-2}
\end{align}  For $\psi_{\text{so}}\ne2n\pi$, the line shape of the
transmission may exhibit two peaks or a single peak profile,
depending on the relations between $\psi_{\text{so}}$ and the
nonzero value of $\delta{E}$.  From
Eq.~(\ref{transmission_plus_minus}), we find that if
\begin{align}
\bar{\gamma}(\psi_{\text{so}})&\equiv4\left( \gamma
_{\psi_{\text{so}} }^{+}\gamma _{\psi_{\text{so}} }^{-}\right)
^{2}\cos ^{2}\left( \psi_{\text{so}} /2\right)\notag\\& -\left[
\left( \gamma _{\psi_{\text{so}} }^{+}\right) ^{2}+\left( \gamma
_{\psi_{\text{so}} }^{-}\right) ^{2}\right] \delta{E} ^{2}\sin
^{2}\left( \psi_{\text{so}} /2\right)
>0, \label{criteria1}
\end{align} is satisfied, then there emerges a profile with two
peaks, as exemplified by the second most front plot on the right
panel of Fig.~\ref{fig2-1}.  The separation between these two peaks
is given by
\begin{align}
\Delta{\omega}&=\Bigg\{ \frac{-\frac{\delta E ^{2}}{2}\sin
^{2}\left( \psi_{\text{so}} /2\right)}{2\cos ^{2}\left(
\psi_{\text{so}} /2\right) }+\notag\\&\frac{\sqrt{\left[
\frac{\delta E ^{2}}{2}\sin ^{2}\left( \psi_{\text{so}} /2\right)
\right]
^{2}+\bar{\gamma}(\psi_{\text{so}})\cos^{2}(\psi_{\text{so}}/2)
}}{2\cos^{2}(\psi_{\text{so}}/2)} \Bigg\}^{1/2}.
\end{align}  When the condition, Eq.~(\ref{criteria1}), is not
fulfilled, the transmission line shape has a single peak, but it is not
a Lorentzian profile (see the plots on the right panel behind the
second most front one in Fig.~\ref{fig2-1}).  The single peak occurs
at $\omega=0$ with the height
\begin{align}
T^{0}\left( \psi_{\text{so}}\neq 2n\pi ,\omega
=0\right)=\frac{16\Gamma _{L}\Gamma
_{R}\delta E ^{2}\sin ^{2}\left( \psi_{\text{so}}/2\right) }{%
\left[ \delta E ^{2}+4\Gamma _{L}\Gamma _{R}\sin ^{2}\left(
\psi_{\text{so}}/2\right) \right] ^{2}}.\label{height_dene0}
\end{align}
The height of the peak depends on $\delta E$ and on $\psi_{\rm so}$. In particular, when $\delta E=0$ then
Eq.~(\ref{lorentz-lineshape}) yields the height $4\Gamma_{L}\Gamma_{R}/\Gamma^2$.

Therefore by observing the zero bias
conductance profile at given electric field without applying the
magnetic flux, one can extract the SOI-induced phase
$\psi_{\text{so}}$ from
Eqs.~(\ref{peak-distance}-\ref{height_dene0}) at that electric
field.  The other way around, one can also fix the electric field at
which $\psi_{\text{so}}=2n\pi$, and find out the value of $\phi$ at
a given magnetic field through a similar procedure by the property,
$T\left(\phi,\psi_{\text{so}}=2n\pi,\omega
\right)=2T^{0}(\phi,\omega)$.  Using these results, one can map out
the dependence of $T^{0}(\varphi,\omega)$ on the effective flux
$\varphi$.  Together with the knowledge on how $\phi$ and
$\psi_{\text{so}}$ depend on the directly tunable magnetic and
electric fields, the transmission for polarized current, namely,
$T^{0}(\varphi^{}_{+},\omega)-T^{0}(\varphi^{}_{-},\omega)$, at given
electric and magnetic fields can be found.  In particular, when the
full polarization conditions are met, the total transmission shall
satisfy
$T(\phi,\psi_{\text{so}},\omega)|_{\varphi^{}_\mp=(2n-1)\pi}=T^{0}(\varphi^{}_\pm,\omega)$.


\section{Dynamics of spin-dependent transport}

\label{sec_DynsSpinDepTrpt}

In this section, we discuss the time evolutions of the spin-dependent
transport.  In the first subsection we study the real
time evolution towards a full spin-polarization in the currents, when
the two conditions given above are fulfilled.  In the second
subsection, we turn to the more general situation to investigate the
interplay between the AB\ and AC interference effects on the
dynamics of spin currents.  In the
third subsection, we discuss how one can obtain the spin-resolved
currents from the total charge currents, using  approaches similar to those
mentioned in a previous section.

\subsection{Time evolution of full\ spin polarization in currents%
}

\label{subsec_DynsSpinDepTrpt_TMFullSpinPolar}

We have discussed the requirements for generating polarized currents
in the steady-state.  However, even when those requirements are
satisfied, the currents during the transient processes for both of
the orthogonal spinors are generally nonvanishing.  Using Eq. (\ref{lead-spin-dep-crnt_form2}), we
explicitly study how the currents for opposite spins along the
characteristic directions in each of the electrodes, $I_{\alpha
\mathbf{\hat{n}_{\alpha};- }}(t)$ and $I_{\alpha
\mathbf{\hat{n}_{\alpha};+ }}(t)$, change in time.

The real-time polarization process is mainly manifested through the
evolution of the current $I_{\alpha \mathbf{\hat{n}_{\alpha};-
}}(t)$, that will eventually decay to zero.
The requirement
Eq.~(\ref{cond-spinpolarcrnt}) with Eq.~(\ref{crnt_alpha_plusminus})
and Eq. (\ref{lead-spin-dep-crnt_form2}) leads to
\begin{align}
&I_{\alpha \mathbf{\hat{n}}_{\alpha};- }(t)=4\Gamma _{\alpha }\times\notag \\
&\int \frac{d\omega }{2\pi }f_{\alpha }\left( \omega \right)
\frac{e^{-\Gamma _{\alpha }t}\left( \Gamma _{\alpha }\cos \omega
t+\omega \sin \omega t\right) -\Gamma _{\alpha }e^{-2\Gamma _{\alpha }t}}{%
\omega ^{2}+\Gamma _{\alpha }^{2}}\ .\label{I_t_minus_alpha}
\end{align}
According to the analysis in Sec.
\ref{subsec_realtimeDQDABSOI_riseSpinTrspt}, under the condition
Eq.~(\ref{cond-spinpolarcrnt}), the corresponding spinless
interferometer, whose current is related to the current carrying the
spinor $\vert\mathbf{\hat{n}}_{\alpha};-\rangle$, has a disconnected
effective configuration as that depicted in Fig.~\ref{fig2}(a). The
result that Eq.~(\ref{I_t_minus_alpha}), as the current for spinor
$|\mathbf{\hat{n}}_{\alpha};- \rangle$ on lead $\alpha$, is not
affected by anything from the other lead $\bar{\alpha}$, verifies
the conclusion in Sec. \ref{subsec_realtimeDQDABSOI_riseSpinTrspt}.

In contrast to the decay of the currents $I_{\alpha \mathbf{\hat{n}_{\alpha};- }}(t)$ toward zero, the time evolution of the currents $I_{\alpha\mathbf{\hat{n}}_{\alpha}\mathbf{;+}%
}\left( t\right) $ generally depends on parameters from both of the
leads.  At the optimal point $\varphi^{} _{+}=2m\pi $ leading to
$\Gamma _{\varphi^{} _{+}}=\Gamma $,
a similar substitution as used for obtaining
Eq.~(\ref{I_t_minus_alpha}) results in,
\begin{eqnarray}
&&I_{\alpha
\mathbf{\hat{n}_{\alpha};+ }}(t) \notag \\
&=&4\Gamma _{\alpha }\int \frac{d\omega }{2\pi }\left\{ f_{\alpha
}\left( \omega \right) \frac{\Gamma +\left( \omega \sin \left(
\omega t\right) -\Gamma \cos \left( \omega t\right) \right)
e^{-\Gamma t}}{\omega
^{2}+\Gamma ^{2}}\right.  \notag \\
&&\left. -\sum_{\alpha ^{\prime }=L,R}\Gamma _{\alpha ^{\prime
}}f_{\alpha ^{\prime }}\left( \omega \right) \frac{\left(
1+e^{-2\Gamma t}-2\cos \left( \omega t\right) e^{-\Gamma t}\right)
}{\omega ^{2}+\Gamma ^{2}}\right\} .
\notag \\
&&  \label{I_t_plus_alpha}
\end{eqnarray}
For this part, the corresponding effective spinless configuration
discussed in Sec. \ref{subsec_realtimeDQDABSOI_riseSpinTrspt} is a
single-level dot coupled to two reservoirs, as shown in
Fig.~\ref{fig2}(b). From Eq.~(\ref{I_t_plus_alpha}), one can also
see that $I_{\alpha
\mathbf{\hat{n}_{\alpha};+ }}(t\rightarrow\infty)\ne0$ with nonzero
bias, as expected from previous discussions.

Equation~(\ref{I_t_minus_alpha}) indicates that the time for
$I_{\alpha \mathbf{\hat{n}_{\alpha};- }}(t)$ to reach its
steady-state value is mainly determined by the term
$e^{-\Gamma_{\alpha}t}$.  The smaller $\Gamma_{\alpha}$ is, the
slower the full spin-polarization of the current in lead $\alpha$ is
reached.  On the other hand, one finds from
Eq.~(\ref{I_t_plus_alpha}) that the time for $I_{\alpha
\mathbf{\hat{n}_{\alpha};+ }}(t)$ to reach its steady-state value is
dominated by the term $e^{-\Gamma t}$.  Therefore it is insensitive
to the specific values taken by the individual couplings
$\Gamma_{L}$ and $\Gamma_{R}$ for a fixed
$\Gamma=\Gamma_{L}+\Gamma_{R}$.  However, the coupling geometry
still affects the magnitude of the full spin-polarized current.  In
the steady states,  Eqs.
(\ref{steady-alpha-plus-crnt},\ref{polarized_trps_steady}) show that
the spin-polarized current can be enhanced by having larger value of
$\Gamma_{L}\Gamma_{R}$.  Note that Eq.
(\ref{steady-alpha-plus-crnt}) is invariant under the exchange of
the couplings, $\Gamma_{L}\leftrightarrow\Gamma_{R}$.

\begin{figure}[h]
\includegraphics[width=8.0cm,
height=9.0cm]{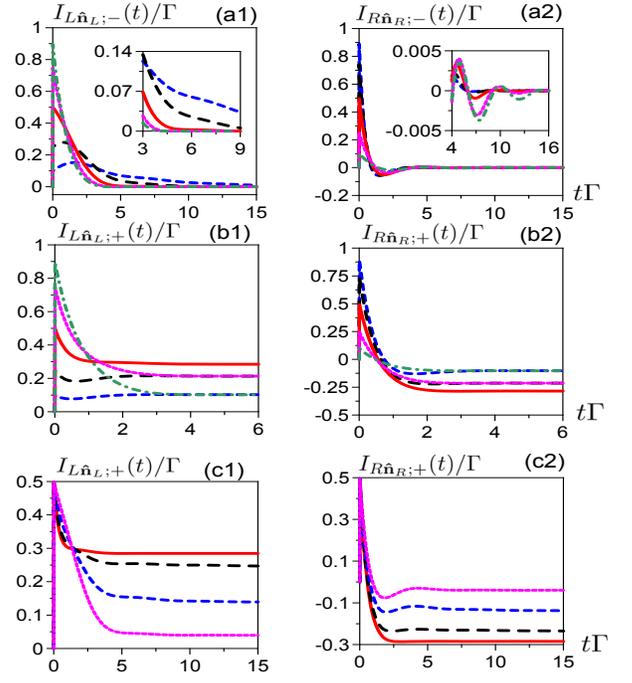} \caption{(color online) Time evolutions of
the currents $I_{\alpha \mathbf{\hat{n}_{\alpha};- }}(t)$ and
$I_{\alpha \mathbf{\hat{n}_{\alpha};+ }}(t)$ under the conditions of
full spin polarizations.  In (a1),(a2),(b1),(b2), the blue dashed
lines are for $(\Gamma_{L},\Gamma_{R})=(0.1,0.9)\Gamma$, the black
long-dashed lines are for
$(\Gamma_{L},\Gamma_{R})=(0.25,0.75)\Gamma$, the red solid lines are
for $(\Gamma_{L},\Gamma_{R})=(0.5,0.5)\Gamma$, the magenta
short-dashed lines are for
$(\Gamma_{L},\Gamma_{R})=(0.75,0.25)\Gamma$ and the green dash-dot
lines are for $(\Gamma_{L},\Gamma_{R})=(0.9,0.1)\Gamma$. In plots
(a1) and (a2), $I_{L\mathbf{\hat{n}}_{L};-}(t)$ and $I_{R
\mathbf{\hat{n}}_{R};- }(t)$ are shown for various coupling
geometries respectively.  The plots (b1) and (b2) respectively are
for $I_{L\mathbf{\hat{n}}_{L};+}(t)$ and $I_{R
\mathbf{\hat{n}}_{R};+ }(t)$ at the optimal point
$\varphi^{}_{+}=2\pi$.  In (c1) and (c2), the time evolutions of the
currents $I_{L\mathbf{\hat{n}}_{L};+}(t)$ and $I_{R
\mathbf{\hat{n}}_{R};+ }(t)$ are plotted for various
$\varphi^{}_{+}$ with a fixed coupling geometry
$(\Gamma_{L},\Gamma_{R})=(0.5,0.5)\Gamma$.  The red solid lines are
with $\varphi^{}_{+}=2\pi$, black long-dashed lines are with
$\varphi^{}_{+}=2\frac{1}{4}\pi$, blue dashed lines are with
$\varphi^{}_{+}=2\frac{1}{2}\pi$ and the magenta short-dashed lines
are with $\varphi^{}_{+}=2\frac{3}{4}\pi$.  In all plots, we have
set $\protect\delta{E}=0$, as one of the
polarization conditions, and a bias $\protect\mu_{L}=-\protect\mu%
_{R}=1.25\Gamma$ is applied at temperature $k_{B}T=\Gamma/20$. This
set of bias and temperature is also assumed in the following
figures.  The instantaneous rising
of the currents to finite values immediately after $t=0$ in these
plots is a direct consequence of the wide-band
limit.\cite{wide-band}} \label{fig3}
\end{figure}

In Fig.~\ref{fig3}(a1,a2) and (b1,b2), we demonstrate the effects of
the coupling geometry discussed above, specified by different values
of $(\Gamma_{L},\Gamma_{R})$, on the time evolution of the
spin-resolved currents.  The curves in these four plots with the
same line styles are with the same pair of couplings
$(\Gamma_{L},\Gamma_{R})$, subject to
$\Gamma_{L}+\Gamma_{R}=\Gamma$.  In Fig.~\ref{fig3} (a1) and (a2),
we illustrate the time evolution of $I_{\alpha
\mathbf{\hat{n}}_{\alpha};- }(t)$.  It shows that a smaller
$\Gamma_{\alpha}$ leads to a slower decay of the current $I_{\alpha
\mathbf{\hat{n}}_{\alpha};- }(t)$, thus a slower process of
spin-polarization (see the insets for a clearer view).  Since we
have set $\mu_{L}>\mu_{R}$, more electrons are involved in the left
than in the right lead.  At later times, this results in generally bigger
magnitudes of currents in the left (comparing the magnitudes in the
insets of Fig.~\ref{fig3}(a1) and (a2).)

In Fig.~\ref{fig3} (b1,b2), the time evolutions of $I_{\alpha
\mathbf{\hat{n}}_{\alpha};+}(t)$ with $\varphi^{}_{+}=2m\pi$ are
inspected.  It shows that different coupling geometries result in
similar times for $I_{\alpha\mathbf{\hat{n}_{\alpha};+}}(t)$ to
reach the corresponding steady-state values. The merging of the
curves (blue medium-dashed line merged with green dash-dotted line
and black long-dashed line merged with magenta short-dashed line)
occur after a time of about $2.5\Gamma^{-1}$, reaching steady-state
values proportional to $\Gamma_{L}\Gamma_{R}$.  The maximized
spin-polarized current is found with
$\Gamma_{L}=\Gamma_{R}=\Gamma/2$.  Comparing Fig.~\ref{fig3} (b)
with Fig.~\ref{fig3} (a) insets, one finds that $I_{\alpha
\mathbf{\hat{n}}_{\alpha};+}(t)$ reaches a stable value generally
faster than the full spin-polarization is arrived.  This is because
the rate for the former, $\Gamma=\Gamma_{L}+\Gamma_{R}$, as a sum of
two couplings, is larger than the rate for the latter, $\Gamma_{L}$
or $\Gamma_{R}$.  The different dependencies of the dynamical
processes of the currents $I_{\alpha
\mathbf{\hat{n}}_{\alpha};-}(t)$ and $I_{\alpha
\mathbf{\hat{n}}_{\alpha};+}(t)$ on the couplings to the reservoirs,
discussed in Sec. \ref{subsec_realtimeDQDABSOI_riseSpinTrspt}, is
then illustrated here.

We further investigate the behavior of $I_{\alpha
\mathbf{\hat{n}_{\alpha}};+}(t)$ when the system is set away from
the optimal point $\varphi^{}_{+}=2m\pi$.  In
Fig.~\ref{fig3}(c1,c2), the time evolution of $I_{\alpha
\mathbf{\hat{n}_{\alpha}};+}(t)$ with different effective fluxes
$\varphi^{}_{+}$ are plotted.  When $\varphi^{}_{+}$ is placed away
from $2m\pi$ toward $(2m+1)\pi$, the steady-state value of this
current, proportional to $\cos^{2}(\varphi^{}_{+}/2)$, as inspected
from Eq. (\ref{steady-alpha-plus-crnt}), decreases.
Figure~\ref{fig3}(c1,c2) also shows that the value of
$\varphi^{}_{+}$ does not obviously affect the time to approach the
steady state but it influences the overall magnitudes of $I_{\alpha
\mathbf{\hat{n}_{\alpha}};+}(t)$ throughout the time evolution. Note
that different values of $\varphi^{}_{+}$ are realized by applying
different magnetic fluxes and SOI parameters.  As long as the
choices of the magnetic fluxes and SOI parameters are subjected to
Eq. (\ref{phi_condi}), the time dependence of $I_{\alpha
\mathbf{\hat{n}_{\alpha}};-}(t)$ remains the same as described by
Eq. (\ref{I_t_minus_alpha}).

\subsection{Dynamics of spin currents}
\label{subsec_DynsSpinDepTrpt_SpinCrnts}

In the last subsection, we have concentrated on the dynamics under
the conditions of reaching fully polarized currents in the
steady-state limit.  In general when the system  deviates from
these conditions, both the spin-up and the spin-down components of
the currents are nonzero and full spin-polarized currents are not
attained.  In this case, instead of studying separately the currents
for spin up and spin down in some specific direction, it is more
interesting to simply focus on the spin currents $\mathbf{I}_{\alpha
}^{\text{S}}\left( t\right)$.

\subsubsection{Spin currents due to SOI without the magnetic flux}
\label{subsubsec_DynsSpinDepTrpt_SpinCrnts_NOSOI}

We first consider the situation with no applied magnetic flux,
namely, $\phi =0$. The effective fluxes are then\ given by
$\varphi^{} _{\pm }=\pm \psi _{\text{so}}$.  In this case,
Eq.~(\ref{spin-crnt-alpha-explicit}) becomes
(upon the use of Eqs.~(19,21) with
the aid of $\frac{d}{dt}N_{0}\left( \varphi^{} _{\pm
},t\right)=I_{L}^{0}\left( \varphi^{}_{\pm},t\right)+I_{R}^{0}\left(
\varphi^{}_{\pm},t\right)$ in Ref.~[\onlinecite{Tu12115453}] through
the identity Eq.~(\ref{lead-spin-dep-crnt_form2})),
\begin{align}
&\mathbf{I}_{\alpha }^{\text{S}}\left(
t\right)\cdot\mathbf{\hat{n}}_{\alpha}=-2\Gamma _{L}\Gamma _{R}\delta E\sin \left( \psi _{\text{so}}\right)\times\notag\\& \int \frac{%
d\omega }{2\pi }\Bigg\{f_{-}\left( \omega \right)\frac{1}{\left\vert \Gamma _{\psi _{\text{so}%
}}\right\vert ^{2}} \frac{d}{dt}\left[ \left\vert u_{p}\left(
t,\omega \right) \right\vert ^{2}\right]\notag\\ &\mp f_{+}\left(
\omega \right) \left[ \text{Re}\left(
\frac{u_{0}^{\ast }\left( t,\omega \right) u_{p}\left( t,\omega \right) }{%
\Gamma _{\psi _{\text{so}}}}\right)  -\Gamma \left\vert
\frac{u_{p}\left( t,\omega \right) }{\Gamma _{\psi
_{\text{so}}}}\right\vert ^{2}\right]  \bigg\}. \label{crnt0-diff}
\end{align}
where $f_{\pm }\left( \omega \right) =f_{L}\left( \omega \right) \pm
f_{R}\left( \omega \right) $. Here
$u_{0,p}\left( t,\omega \right) =\int_{0}^{t}d\tau e^{i\omega \tau
}u_{0,p}\left( \psi _{\text{so}},\tau \right) $ where $u_{0,p}\left(
\psi _{\text{so}},\tau \right)$ is
equal to $u_{0,p}(\tau)$ of Eq.~(16) in
Ref.~[\onlinecite{Tu12115453}] with $\phi$ there replaced by $\psi
_{\text{so}}$. The upper sign is for $\alpha=L$ and the lower sign
is for $\alpha=R$. Equation (\ref{crnt0-diff}) shows that if we set
$\delta E=0$, then the spin currents remain zero for all time $t$.
To generate a non-vanishing spin current, one has to lift up the
degeneracy. From
Ref.~[\onlinecite{Tu12115453}],
one finds that $\delta{E}=0$ leads to $I_{\alpha}^{0}\left( \psi _{%
\text{so}},t\right) =I_{\alpha}^{0}\left( -\psi
_{\text{so}},t\right) $ for $\alpha=L,~R$ and phase rigidity is
kept for the underlying spinless system  throughout the time.
Therefore, generating spin currents by lifting up the degeneracy is
equivalent to the temporary breaking of phase rigidity in the
spinless DQD interferometer, as pointed out in Ref.
[\onlinecite{Tu12115453}]. Besides the energy splitting,
Eq.~(\ref{crnt0-diff}) also explicitly reveals the necessity of the
presence of SOI for the existence of the spin currents, through the
term $\sin\left(\psi_{\text{so}}\right)$.  When SOI is absent,
$\psi_{\text{so}}=0$, then $\sin\left(\psi_{\text{so}}\right)=0$,
directly leading to $\mathbf{I}_{\alpha }^{\text{S}}\left(
t\right)=0$. Eq.~(\ref{crnt0-diff}) also shows that the spin
currents $\mathbf{I}_{\alpha }^{\text{S}}\left( t\right)$ will
approach zero at long times. The non-vanishing spin currents can
thus only be sustained transiently.

\begin{figure}[h]
\includegraphics[width=8.5cm,
height=7.0cm]{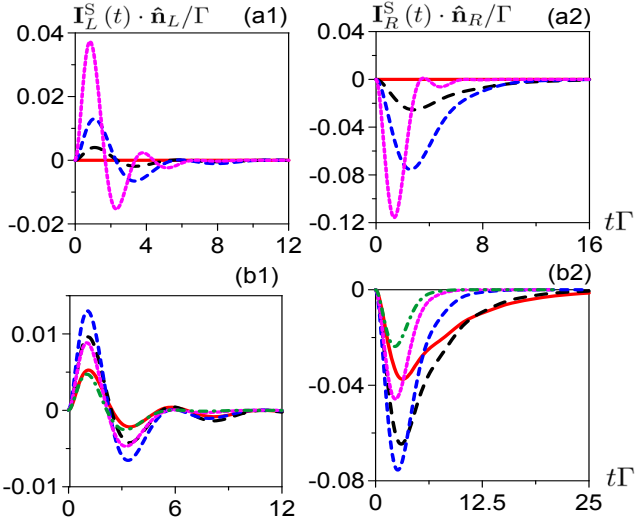} \caption{(color online) Time evolutions of
the spin currents.  The couplings are set to
$\Gamma_{L}=\Gamma_{R}=\Gamma/2$ here. In plots (a1) and (a2) all
curves are with $\psi_{\text{so}}=0.5\pi$.  The red solid lines are
for $\protect\delta{E}=0$, the black long-dashed lines are for
$\protect\delta{E}=0.15\Gamma$, the blue dashed lines are for
$\protect\delta{E}=0.5\Gamma$ and the magenta short-dashed lines are
for $\protect\delta{E}=2\Gamma$.  In plots (b1) and (b2) we fix
$\delta{E}=0.5\Gamma$.  The red solid lines are for
$\psi_{\text{so}}=0.125\pi$, the black long-dashed lines are for
$\psi_{\text{so}}=0.25\pi$, the blue dashed lines are for
$\psi_{\text{so}}=0.5\pi$, the magenta short-dashed lines are for
$\psi_{\text{so}}=0.75\pi$ and the green dash-dot lines are for
$\psi_{\text{so}}=0.875\pi$. }\label{fig4}
\end{figure}

The above discussions show that the magnitudes of the transient spin
currents are mainly determined by $\delta E$ and $\sin\left(\psi
_{\text{so}}\right)$.  In Fig.~\ref{fig4} (a1, a2), we study the
effects of various $\delta E$'s on the time evolutions of the spin
currents.  The results show that splitting the degeneracy generally
enhances the transient spin flow (compare the curves for smaller and
bigger $\delta E$), as indicated by Eq.~(\ref{crnt0-diff}).  Since
bigger $\delta E$ also implies faster relaxation, we observe a
shorter span of nonzero spin flow with bigger energy splitting.  The
dependencies of the spin currents on SOI, through the SOI-induced
phase $\psi _{\text{so}}$, are presented in  Fig.~\ref{fig4} (b1,
b2).  The transient magnitudes increase with increasing values of
$\sin\left(\psi _{\text{so}}\right)$.

\subsubsection{Spin currents due to SOI with the magnetic flux}
\label{subsubsec_DynsSpinDepTrpt_SpinCrnts_withSOI}

The above discussions have shown that it is not possible to generate
spin currents at degeneracy purely by the act of SOI.  Even when
the degeneracy between the on-site energies of the DQD is lifted up, the
spin currents only survive transiently.  The magnetic flux is thus
indispensable to sustain non-vanishing spin currents in the long time
limit.

\begin{figure}[h]
\includegraphics[width=9.0cm,
height=7.0cm]{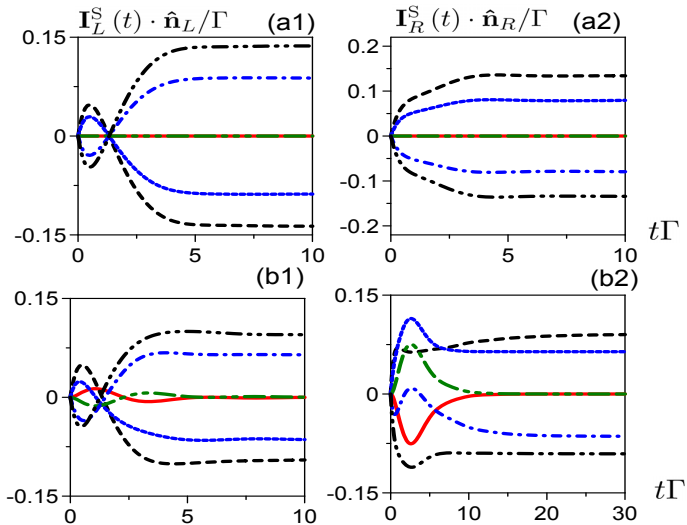} \caption{(color online) Various magnetic
flux values are tested for fixed SOI strength. In all plots we let
$\psi_{\text{so}}=0.5\pi$.  In plots (a1) and (a2), we set
$\protect\delta{E}=0$.  The red solid lines for $\protect\phi =0$
and the dark green dash-dash-dot-dot-dot lines for
$\protect\phi=\protect\pi$ are overlapped.  The black-dashed lines
for $\protect\phi =0.4\protect\pi $ and the black dash-dot-dot lines
for $\protect\phi=1.6\protect\pi$ differ by a sign. The blue
short-dashed lines for $\protect\phi =0.8\protect\pi $ and the blue
dash-dot lines for $\protect\phi =1.2\protect\pi $ also differ by a
sign.  In plots (b1) and (b2), the degeneracy is lifted as
$\protect\delta {E}=0.5\Gamma$ and the same choices of $\phi$ used
in (a1) and (a2) are used also here. }\label{fig5}
\end{figure}

The versatility of the combination of the AB effect and the SOI for
attaining various spin currents is demonstrated in Fig.~\ref{fig5}
for both  $\delta E=0 $ and $\delta E\neq 0$. \ The values of $\phi
$ are so\ chosen that one can attain various distinct results for
the spin current.\ \ In Fig.~\ref{fig5}(a1,a2), we illustrate that
at $\delta E=0$, spin currents can be generated and sustained in the
steady states by simultaneously setting $\phi\neq 0$ and
$\sin(\psi_{\text{so}})\neq 0$.  The solution
Eq.~(\ref{spin-crnt-alpha-explicit}) implies that one can reverse
the sign of the spin currents by just adjusting the flux $\phi$,
without altering the SOI parameters leading to the changes of
$\mathbf{\hat{n}}_{\alpha}$ and $\psi_{\text{so}}$.  At degeneracy,
the currents $I^{0}_{\alpha}(\varphi^{}_{\pm},t)$ for the effective
spinless system depend on the effective flux $\varphi^{}_{\pm}$ only
through the term $\cos\varphi^{}_{\pm}$ (see
Ref.~[\onlinecite{Tu12115453}]).
Therefore, for all times $t$, the sign of $\mathbf{I}_{\alpha
}^{\text{S}}\left(
t\right)\cdot\mathbf{\hat{n}}_{\alpha}|_{\phi=n\pi+\Delta \phi}$ and
that of $\mathbf{I}_{\alpha }^{\text{S}}\left(
t\right)\cdot\mathbf{\hat{n}}_{\alpha}|_{\phi=n\pi-\Delta \phi}$, for $n$ being an arbitrary integer and $%
\Delta \phi $ being nonzero, are opposite to each other, as
indicated in Fig.~\ref{fig5}(a1,a2).  By lifting up the degeneracy,
the dependence of $I^{0}_{\alpha}(\varphi^{}_{\pm},t)$ on
$\varphi^{}_{\pm}$ appears from both of the terms
$\cos\varphi^{}_{\pm}$ and $\sin\varphi^{}_{\pm}$.  As a result, the
spin currents with $\delta{E}\ne0$ are not perfectly antisymmetric
with respect to $\phi=n\pi$, as shown by Fig.~\ref{fig5} (b1,b2).


\subsection{Deduction of the
spin-resolved currents from the spin-independent total charge
current}

\label{subsec_DynsSpinDepTrpt_DeductionSpinCrnt}

In Sec. \ref{subsec_SteadyState_ExtractSpTrsm}, we have discussed
how one can extract the values of $\psi_{\text{so}}$ and $\phi$ at
given electric and magnetic fields.  With this knowledge in mind, by
applying similar procedures, we can also obtain the magnitude of the
spin current, $\mathbf{I}_{\alpha }^{\text{S}}\left(
t\right)\cdot\mathbf{\hat{n}}_{\alpha}$, from the values of the
spin-independent total charge current.  Explicitly, the
spin-independent total charge current, Eq.~(\ref{totcrnt}), at given
electric and magnetic fields is
\begin{align}
I_{\alpha }(\phi,\psi_{\text{so}},t)=I_{\alpha }^{0}\left(
\varphi^{}_{+} ,t\right)+I_{\alpha }^{0}\left( \varphi^{}_{-}
,t\right).\label{spinless-crnt-decompose1}
\end{align} From Eqs.~(19,21) in Ref.~[\onlinecite{Tu12115453}], the current on lead $\alpha$ for the effective
spinless DQD system with flux $\varphi$ can be split into two terms,
\begin{subequations}
\label{spinless-crnt-decompose2}
\begin{align}
I_{\alpha }^{0}\left( \varphi ,t\right) =\bar{I}_{\alpha }^{0}\left(
\varphi ,t\right) +\delta E \sin \left( \varphi \right)
\bar{I}_{\alpha }^{\gamma }\left( \varphi
,t\right),\label{spinless-crnt-decompose2-1}
\end{align} where  $\bar{I}_{\alpha }^{0}\left(
\varphi ,t\right)$ and $\bar{I}_{\alpha }^{\gamma }\left( \varphi
,t\right)$ satisfy,
\begin{align}
\bar{I}_{\alpha }^{0}\left(\varphi ,t\right)=\bar{I}_{\alpha
}^{0}\left(-\varphi ,t\right),~\bar{I}_{\alpha }^{\gamma }\left(
\varphi ,t\right)=\bar{I}_{\alpha }^{\gamma }\left(-\varphi
,t\right).\label{spinless-crnt-decompose2-2}
\end{align}
\end{subequations}

Setting zero magnetic field $\phi=0$ in
Eq.~(\ref{spinless-crnt-decompose1}) with the property given by
Eq.~(\ref{spinless-crnt-decompose2}), the dependence of the total
charge current on $\psi_{\text{so}}$ becomes
\begin{align}
I_{\alpha }(\phi=0,\psi_{\text{so}},t)=2\bar{I}_{\alpha }^{0}\left(
\psi_{\text{so}} ,t\right).
\end{align}  The dependence of $\bar{I}_{\alpha }^{0}\left(
\psi_{\text{so}} ,t\right)$ on various $\psi_{\text{so}}$'s can thus
be found from the total current $I_{\alpha
}(\phi=0,\psi_{\text{so}},t)$ under different applied electric
fields with zero magnetic flux.  The dependence of $\bar{I}_{\alpha
}^{\gamma }\left( \varphi ,t\right)$ on $\varphi$ can be found
through the following approach.  By fixing the electric field at
$\psi_{\text{so}}=2n\pi$, this part of the current is related to the
total current and to the part that is already known, $\bar{I}_{\alpha
}^{0}\left( \varphi ,t\right)$, via the relation
\begin{align}
2\delta E \sin \left( \phi \right) \bar{I}_{\alpha }^{\gamma }\left(
\phi ,t\right) =I_{\alpha }\left( \phi ,\psi _{\text{so}}=2n\pi
,t\right) -2\bar{I}_{\alpha }^{0}\left( \phi ,t\right).
\end{align}  The values of $\bar{I}_{\alpha
}^{\gamma}\left( \phi ,t\right)$ for different $\phi$'s can thus be
deduced from the total current and $\bar{I}_{\alpha }^{0}\left( \phi
,t\right)$ by applying the corresponding  magnetic fields.
Knowing $I^{0}_{\alpha}(\varphi,t)$ at various effective fluxes
$\varphi$, one can deduce the spin-resolved current by the
virtue of Eq.~(\ref{lead-spin-dep-crnt_form2}).


\section{summary and conclusion}
\label{sec_conclusion}

In this paper, we have explored the real-time dynamics of
spin-dependent electron transport through a DQD AB interferometer
with SOI. We have obtained the real-time evolution of the
spin-resolved currents, Eq.~(\ref{rewrite-spin-crnt}), and the
subsequent spin currents Eq.~(\ref{spin-crnt-alpha-explicit}). These
expressions fully describe the dynamical evolution of the
spin-polarizations in the electron transport from initially
completely unpolarized interferometers. We have particularly
investigated the real-time evolution of the currents towards fully
spin-polarized transport. We have also explored the interplay
between the SOI and AB interferences in the dynamics of spin flows.
Out of these studies, we draw the following conclusions.

(1) The effects of SOI on the transport currents are attributed to
the SOI-induced phase $\psi_{\text{so}}$ and the characteristic
directions, $\mathbf{\hat{n}}_{L}$ and $\mathbf{\hat{n}}_{R}$. They
are fully determined respectively as the eigenvalue and eigenvectors
of the unitary spin rotations around the loop. Spin-polarizations of
currents in each lead are only developed along their characteristic
directions throughout all times. In general,
$\mathbf{\hat{n}}_{L}\ne\mathbf{\hat{n}}_{R}$.

(2) The currents carrying electrons with spins along the
characteristic directions are equal
to the currents of the effective spinless system with the flux
replaced by the effective fluxes, as described by
Eq.~(\ref{lead-spin-dep-crnt_form2}). Such connections explicitly
reveal that it is the difference between the effective fluxes,
$\varphi_{\pm}=\phi\pm\psi_{\text{so}}$, caused by the SOI, that
gives rise to the spin-polarizations.

(3) When fully polarized currents for spin-up electrons occur, the
effective spinless system underlying the spin-down current
corresponds to a disconnected configuration, as shown in
Fig.~\ref{fig2}(a). Therefore, the decay dynamics of spin-down
current in one lead is independent of that in the other lead. In
contrast, the dynamics of building spin-up current in one electrode
is correlated with that in the other electrode. This essential
picture can serve to discern the occurrence of full
spin-polarization. One could test it by monitoring the spin-resolved
currents in one side while changing parameters of the other side.


(4) The realization of full spin polarization has received
considerable attention, as it supplies  spin-polarized electron
sources and allows the manipulation of flying spin qubits. Therefore
it is important to know how to implement fast full spin-polarization
and attain the polarized currents of tunable magnitudes. At full
polarization condition, we found that increasing the coupling
strength to the leads effectively increases the pace toward full
spin-polarization. As an interferometer, the magnitude of the
resulting currents is largely controlled by the interference phase,
which is determined here by the effective fluxes. They are tunable
via the magnetic flux and the electric field, affecting the
SOI-induced phase.


(5) The connection between the spin-polarized currents and the
currents of the effective spinless system provides the underlying
physical picture for the working of the spin transport in this DQD
AB interferometer. Due to the phase rigidity of the effective
spinless system as a two-terminal setup, SOI alone cannot give rise
to steady-state spin currents. The indispensable role of the
magnetic flux in maintaining nonzero spin currents to the
steady-state limit reflects the essence of the interplay between
charge and spin interferences. This can be tested by comparing the
steady-state spin currents with and without the applied magnetic
flux.

(6) Spin-independent total charge currents are readily measurable in
experiments. We have shown how one can extract currents of
characteristic spins in this system from measurements of the total
charge currents at properly chosen electric and magnetic fields.
Such measurements can thus be used for testing the properties of the
spin-resolved transport concluded above.

The model we considered in this paper could be constructed from
gate-defined QDs made of materials of high carrier density,
providing large screening of Coulomb interactions such that
electrons are effectively noninteracting. With sufficient orbital
level spacing and applying a bias less than such spacing, one can
make only a single orbital in each dot participate in the transport.
The QDs could be connected to electron reservoirs via gated quantum
wires to implement the SOI tunable by the gate electric field. An
ongoing experimental issue concerns the possibility to detect
 spin-polarized electrical currents using only electrical
means.\cite{Otsuka09195313,Debray09759,Kim12054706,Chen12177202}
Modulating electron transport via interference in ring-like
structures with the AB effect\cite {Hatano11076801,Verduijn13033020}
and spin interference\cite
{Konig06076804,Bergsten06196803,Nagasawa12086801,Nagasawa132526} are
of much experimental interest. The analysis of the time evolution of
spin-resolved transport  for this DQD interferometer could add a
momentum to this progressing research direction.



\begin{acknowledgements}
 Work at NCKU is partially supported by the National Science Council
(NSC) of the ROC, under Contract No. NSC102-2112-M-006-016-MY3, by the Headquarters of
University Advancement at the National Cheng Kung University, which is sponsored by the Ministry of Education, Taiwan, ROC
and  from the National Center for Theoretical Science
of NSC and the High Performance Computing Facility in the National
Cheng Kung University. Work at Ben Gurion University was supported by grants from the Israel Science Foundation (ISF) and from the U.S.-Israel
Binational Science Foundation (BSF).
\end{acknowledgements}

\appendix

\section{NEGF formalism for time-dependent transport currents}
\label{sec_formalism}

Here we summarize formulations for studying non-equilibrium electron
transport through a class of nano-electronic structures. In order to
consider spin-dependent dynamics, here we label both the charge and
the spin degrees of freedom explicitly. We also assume that electron
reservoirs are free from SOI. The Hamiltonian of the total system is
then given by Eq.~(\ref{general_total_H}), where the Hamiltonian of
the central area is generally
\begin{equation}
\mathcal{H}_{\mathrm{S}}=\sum_{ij,\sigma \sigma ^{\prime
}}E_{i\sigma ,j\sigma ^{\prime }}a_{i\sigma }^{\dag }a^{}_{j\sigma
^{\prime }}\label{general_systemH-A}
\end{equation}%
with $i,j$ labeling orbital states and $\sigma ,\sigma ^{\prime }$
denoting the spins. The Hamiltonian for sum of electron reservoirs
is Eq.~(\ref{general_envH}), with $\alpha$ running over all
considered electrodes. The tunneling between the central system and
the leads is described by Eq.~(\ref{general_tulH}). In all the above
equations, the spins are quantized along an arbitrary direction.


The spin-resolved transient current at time $t$ is defined by,
\begin{subequations}
\label{def-spinresolvcrnt_asdt}
\begin{align}
I_{\alpha \sigma
}(t)=-\frac{d}{dt}\mathrm{tr}_{\text{tot}}[\mathcal{N}_{\alpha,\sigma}\rho
_{\text{tot}}(t)],
\end{align} where
\begin{equation}
\mathcal{N}_{\alpha,\sigma}=\sum_{\bm k\in \alpha }c_{\alpha \bm k\sigma%
}^{\dag }c^{}_{\alpha \bm k\sigma}\label{def_spinreslv_particnmbr},
\end{equation}
\end{subequations}
is the total particle number operator for spin $\sigma$ in lead
$\alpha$ and $\rho _{\mathrm{tot}}(t)$ is the total density matrix
of the central system plus the electron reservoirs at time $t$. Here
$\mathrm{tr}_{\text{tot}}$ denotes the trace over the total system.

As usual, we assume\cite{Leggett871} that at the initial time
$t=t_0$, the central system is decoupled from the leads, and the
leads are at thermal equilibrium
with the chemical potential $\mu _{\alpha\sigma }$ and the temperature $%
T_{\alpha\sigma } $ for electron with spin $\sigma$ in lead $\alpha
$, whose Fermi distribution function is given by,
\begin{align} f_{\alpha\sigma
}(\omega )=1/[e^{(\omega -\mu _{\alpha \sigma})/k_{B}T_{\alpha
\sigma}}+1],
\end{align}
where $k_{B}$ is the Boltzmann constant. If the central area
initially contains no excess electrons, then the real-time current
carrying electrons of spin $\sigma$ from lead $\alpha$ in terms of
Keldysh NEGF reads\cite{Jauho945528,Haug08}
\begin{align}
&I_{\alpha \sigma}(t) =-2\mathrm{ReTr}\nonumber\\
&\int_{t_{0}}^{t}d%
\tau \big\{{\bm \Sigma}^{r}_{\alpha \sigma ^{ }}(t,\tau ){\mathbf G}^{<}(\tau,t)+%
{\bm \Sigma}^{<}_{\alpha \sigma ^{ }}(t,\tau
)\mathbf{G}^{a}(\tau,t)\}. \label{kd-crntafa0}
\end{align} Throughout the paper, we use units in which $\hbar =e=1
$. One can also derive the same current formula through a density matrix formalism,\cite{Jin10083013} as used in Sec. \ref{subsec_realtimeDQDABSOI_TimeDepSpinCrntGF}, and the two expressions can be identified via Eqs.(\ref{gtg-gtos},\ref{uvn-gtos}). The retarded and the lesser self-energies are
\begin{subequations}
\label{kd-SE}
\begin{align}
&{\bm \Sigma}^{r}_{\alpha \sigma ^{ }}(t,\tau )=-i\theta(t-\tau)\int \frac{d\omega }{2\pi }{\bm\Gamma }%
^{\alpha \sigma }(\omega )e^{-i\omega (t-\tau) },  \label{kd-rdSE_ldspn} \\
& {\bm \Sigma}^{<}_{\alpha \sigma ^{ }}(t,\tau
)=i\int \frac{d\omega }{2\pi }%
f_{\alpha\sigma }(\omega ){\bm\Gamma }^{\alpha \sigma }(\omega
)e^{-i\omega (t-\tau) }\text{,}  \label{kd-lessSE_ldspn}
\end{align} respectively, with $\theta$ being the step function. They are defined via the level-broadening function,
\begin{equation}
\left[ {\bm\Gamma }^{\alpha \sigma }(\omega )\right]
_{i\sigma',j\sigma''}=2\pi \sum_{\bm {k}\in \alpha
}V_{i\sigma',\alpha \bm{k}\sigma }V_{j\sigma'',\alpha \bm{k}\sigma
}^{\ast }\delta (\omega -\epsilon _{\alpha \bm k}),
\label{lv-brd_func}
\end{equation}
\end{subequations} The retarded and the advanced Green functions are
defined by
\begin{subequations}
\label{kd-rdGF}
\begin{align}
&\left[\mathbf{G}^{r}(t,\tau)\right]_{i\sigma,j\sigma'}=-i\theta(t-\tau)\left\langle\left\{a^{}_{i\sigma}(\tau),a^{\dagger}_{j\sigma'}(t)\right\}\right\rangle,\\
&\left[\mathbf{G}^{a}(\tau,t)\right]_{i\sigma,j\sigma'}=i\theta(t-\tau)\left\langle\left\{a^{}_{i\sigma}(\tau),a^{\dagger}_{j\sigma'}(t)\right\}\right\rangle.
\end{align}
They are related by
$\mathbf{G}^{a}(\tau,t)=\left[\mathbf{G}^{r}(t,\tau)\right]^{\dagger}$.
The retarded Green function follows the equation
\begin{equation}
\left[i\partial_{t}-\boldsymbol{E}\right]\mathbf{G}^{r}(t,\tau)-\int_{\tau}^{t}d\tau'{\bm
\Sigma}^{r}_{}(t,\tau')\mathbf{G}^{r}(\tau',\tau)=\delta(t-\tau),
\end{equation}
\end{subequations} and the lesser Green function is given by
\begin{align}
\label{kd-lessGF} {\mathbf
G}^{<}(\tau,t)=\int_{t_{0}}^{\infty}d\tau'\int_{t_{0}}^{\infty}d\tau''
\mathbf{G}^{r}(\tau,\tau'){\boldsymbol
\Sigma}^{<}(\tau',\tau'')\mathbf{G}^{a}(\tau'',t).
\end{align}
 Here $\bm E$ is the
energy matrix of the central system while
\begin{subequations}
\begin{align}
{\bm \Sigma}^{r}_{}(\tau,\tau' )=\sum_{\alpha\sigma}{\bm
\Sigma}^{r}_{\alpha \sigma ^{ }}(\tau,\tau' ),\label{Kd-reSE}\\
{\bm \Sigma}^{<}_{}(\tau,\tau' )=\sum_{\alpha\sigma}{\bm
\Sigma}^{<}_{\alpha \sigma ^{ }}(\tau,\tau' )\label{Kd-lessSE}
\end{align}
\end{subequations} are sums of individual self-energies induced by
coupling to each of the leads. In the definitions Eqs.
(\ref{kd-rdGF},\ref{kd-lessGF}), the time-dependent field operators
are those in the Heisenberg picture and the bracket denotes the
average over the initial state,
$\langle\cdot\rangle=\text{tr}_{\text{tot}}(~\cdot~\rho_{\text{tot}}(t_{0}))$.

By specifying the level-broadening function, Eq. (\ref%
{lv-brd_func}), and therefore the self-energies, Eq. (\ref{kd-SE}),
one can substitute them into Eqs. (\ref{kd-rdGF}, \ref{kd-lessGF})
for solving the Green functions in the time domain. \ The real time
currents can then be found by substituting these Green functions and
self-energies into Eq. (\ref{kd-crntafa0}).




\begin{thebibliography}{99}
\bibitem{footnote0}
In Ref. [\onlinecite{Aharony11035323}], the authors diagonalized the
quantities, $\mathbf{W}\mathbf{W}^{\dag}$ and
$\mathbf{W}^{\dag}\mathbf{W}$, where
$\mathbf{W}=\gamma_{b}U_{b}+\gamma_{c}U_{c}$ (using the notations in
that paper), in which $\gamma_{b,c}$ are some real scalar
coefficients relating to couplings to the leads.  The unitary spin
rotations are identified to be $U_{b}=U^{L1}U^{1R}$ and
$U_{c}=U^{L2}U^{2R}$. The spin rotations around the loop commute
with these quantities, namely,
$[\mathbf{W}\mathbf{W}^{\dag},U^{L}]=0$ and
$[\mathbf{W}^{\dag}\mathbf{W},U^{R}]=0$.  Therefore the eigenstates
of $\mathbf{W}\mathbf{W}^{\dag}$ ($\mathbf{W}^{\dag}\mathbf{W}$)
obtained there are also those of $U^{L}$ ($U^{R}$) as needed here.
\bibitem{wide-band}The wide-band limit means that there is an infinite
number of levels in the reservoir that couple equally strong to the
discrete levels in the quantum dot.  So every level in the reservoir
within this band of infinite width tries to tunnel to the discrete
levels, causing an immediate finite current when the contact begins.
This feature of wide-band level-broadening function is also
documented.\cite{Jauho945528,Schmidt08235110}  We demonstrate all
numerical results in the wide-band limit here for simplicity and for
staying connected with our previous work\cite{Tu12115453}.
\end{thebibliography}
\end{document}